\documentclass[11pt,sort&compress]{elsarticle}

\input{preamble.tex}

\begin{document}

\hypersetup{
  linkcolor=darkrust,
  citecolor=seagreen,
  urlcolor=darkrust,
  pdfauthor=author,
}

\begin{frontmatter}
    \title{Limits of constant-parameter constitutive models \\ for hydrogels under inertial cavitation}
    
    \author[add1]{Tianyi Chu\corref{cor1}}
    \ead{tchu72@gatech.edu}
    \author[add2]{Joseph Beckett}
    \author[add2]{Zhiren Zhu}
    \author[add2]{Jonathan B.\ Estrada}
    \author[add1,add3,add4]{Spencer~H.~Bryngelson}
 
    \address[add1]{School of Computational Science $\&$ Engineering, Georgia Institute of Technology, Atlanta, GA 30332, USA \vspace{-0.125cm}}
    \address[add2]{Department of Mechanical Engineering, University of Michigan, Ann Arbor, MI 48105, USA \vspace{-0.125cm}}
    \address[add3]{Daniel Guggenheim School of Aerospace Engineering, Georgia Institute of Technology, Atlanta, GA 30332, USA \vspace{-0.125cm}}
    \address[add4]{George W.\ Woodruff School of Mechanical Engineering, Georgia Institute of Technology, Atlanta, GA 30332, USA}
    \cortext[cor1]{Corresponding author}
    \date{}
\end{frontmatter}

\begin{abstract}

Mechanical characterization of soft materials at high strain rates is challenging due to their high compliance, nonlinear viscoelastic behavior, and potentially history-dependent responses.
Inertial microcavitation rheometry (IMR) addresses this challenge by coupling laser-induced cavitation (LIC) experiments with numerical simulations of bubble dynamics models to infer constitutive models and material parameters.
Both IMR and its variants infer parameters that depend on the chosen fitting window, which suggests that a constant-parameter constitutive model is insufficient to describe the full cavitation event.
We use this window dependence to identify when the constant-parameter assumption fails, rather than to report a single effective parameter set.
The constitutive parameters are estimated over moving, overlapping windows using a modified iterative ensemble Kalman smoother with multiple data assimilation (MIEnKS-MDA).
Within the neo-Hookean Kelvin--Voigt (NHKV) constitutive model, we obtain time-resolved estimates of the constitutive response in polyacrylamide (PAAm) hydrogels with different crosslinker concentrations.
The inferred shear modulus and viscosity generally decrease and then plateau during cavitation, while exhibiting relatively weak temperature sensitivity.
For gelatin gels, by contrast, the inferred property evolution shows a pronounced temperature dependence, with distinct trends at low and high temperatures.
Moreover, both the apparent shear modulus and viscosity exhibit significant variations during the first two bubble collapses.
These results show that time-resolved parameter estimation within the prescribed NHKV constitutive structure can diagnose where the constant-parameter model assumption falls short during cavitation, thereby guiding the development of improved physics-based models of complex bubble--material interactions.



\textit{Keywords:} Viscoelasticity; Rheology; Data assimilation; High strain rate; Inertial cavitation; Hydrogels
\end{abstract}

\blfootnote{Code available at \url{https://github.com/InertialMicrocavitationRheometry}}

\section{Introduction}


Soft biological and tissue-like materials can experience extreme loading during blast exposure, shock-wave impact, laser irradiation, and histotripsy, where deformation unfolds on microsecond timescales.
The combination of rapid energy deposition and large strains drives the response into the high-strain-rate regime (\qtyrange{1e3}{1e8}{\per\second}), where these materials exhibit strongly nonlinear viscoelastic behavior that can differ markedly from their quasi-static mechanics.
Characterization of this regime is important to applications ranging from impact biomechanics to focused ultrasound and soft-matter engineering, including tissue phantom studies, laser surgery diagnosis, and targeted cellular interventions such as DNA manipulation~\citep{mancia2019modeling,vlaisavljevich2016visualizing,brujan2006stress,bailey2003physical,brennen2015cavitation}.
Yet, such characterization remains challenging.
The high compliance of soft materials~\citep{arora1999compliance,chen2010split}, coupled with large deformations and constitutive responses beyond linear elasticity~\citep{lin2009spherical,style2013surface}, complicates both modeling and inference.
Experiments are further constrained by limited optical access, sensitivity to boundary conditions, and coupled thermal and diffusive effects during rapid loading.

Cavitation rheometry is an approach for measuring finite-deformation mechanics under high strain rates and at small physical scales.
Focused energy sources, such as lasers, ultrasound, or shock waves, are often used to trigger inertial cavitation in compliant materials, thereby driving rapid, large deformations at high strain rates suitable for constitutive characterization.
In particular, laser-induced cavitation (LIC) has attracted increasing attention because it enables efficient, readily controlled, and repeatable measurements in soft materials.
In a typical LIC experiment, a single green laser pulse is reshaped and focused onto a soft material sample.
In transparent hydrogels, the resulting cavitation event is recorded with an ultra-high-speed camera at frame rates exceeding $\SI{1e6}{\fps}$~\citep{maxwell2013probability,wilson2019comparative,yang2020extracting,liang2022comprehensive}.
The bubble-radius history is then extracted from the image sequence via post-processing and used to quantify the cavitation dynamics.
Building on LIC, \citet{estrada2018high} introduced inertial microcavitation rheometry (IMR) by coupling measured bubble evolutions with physics-based bubble dynamics models.
Using a prescribed constitutive model, IMR infers effective constitutive parameters over a fitting window, yielding an averaged response.
IMR accommodates a range of viscoelastic constitutive models and has been applied to characterize the mechanical behavior of commonly used biomimetic hydrogels, including
polyacrylamide (PAAm)~\citep{estrada2018high,yang2020extracting,buyukozturk2022particle},
agarose~\citep{mancia2021acoustic,yang2022mechanical}, and gelatin~\citep{bremer2024ballistic,sanchez2025hierarchical}. 
Beyond prescribing a constitutive model \emph{a priori}, \citet{sanchez2025hierarchical} recently proposed a Bayesian hierarchical framework for model selection, improving reliability under uncertainty and reducing model-form error.


In practice, standard IMR characterization is challenging as rheometry in high-strain-rate regimes is intrinsically uncertainty-prone: experimental variability, measurement noise, and model-form error all contribute to a non-negligible model-data mismatch.
When posed as a deterministic inverse problem, IMR often requires repeated cavitation measurements, typically on the order of tens of experiments, to reduce experimental variability and obtain robust parameter estimates~\citep{estrada2018high,yang2020extracting,spratt2021characterizing}.
In addition, the strong nonlinearity of bubble dynamics typically demands dedicated optimization strategies to identify best-fit parameters among the material property space, rendering brute-force IMR searches computationally expensive.
To mitigate these limitations, several IMR variants, including reduced-order and data assimilation (DA) versions, have been developed to reduce computational cost and account for experimental uncertainty.

Very recently, \citet{zhu2025parsimonious} introduced parsimonious IMR~(pIMR) as a computationally efficient alternative for IMR by changing the inference target from the full bubble-radius trajectory to a single scalar observable, the bubble collapse time.
Compared to standard IMR, pIMR eliminates the need for forward simulation of the bubble dynamics, enabling rapid exploration of the material-parameter space.
This efficiency comes at the cost of reduced temporal information.
Since pIMR does not use the post-collapse dynamics, it effectively identifies an averaged material response over the first-collapse time window.
Consequently, inference based solely on the first collapse time can introduce more uncertainty, particularly in the presence of measurement noise.
Accurate pIMR characterization, therefore, requires sufficiently reliable time-resolved diagnostics to identify the first-collapse times.
Moreover, pIMR remains a deterministic procedure and does not explicitly model variability in the measured collapse times, which can complicate identification of a unique best-fit estimate.
These considerations motivate statistical approaches that directly incorporate experimental uncertainty.

Data-assimilation (DA) methods provide a systematic means of addressing model-data mismatch by combining theoretical predictions and observations within a probabilistic framework that represents uncertainty in both.
In particular, the ensemble Kalman filter (EnKF) is among the most widely used DA methods due to its straightforward formulation and ease of implementation~\citep{evensen1994sequential}.
Applications span oceanography~\citep{evensen2003ensemble,sakov2012topaz4}, atmospheric science~\citep{houtekamer1998data, burgers1998analysis, whitaker2002ensemble}, and engineering~\citep{ghanem2006health, aanonsen2009ensemble}, and the method has attracted increasing attention in rheology~\citep{brezzi2016new,chen2024multivariate,lyu2025data}.
Several variants of EnKF have also been developed, including the ensemble Kalman
smoother (EnKS)~\citep{evensen2000ensemble}, iterative EnKS (IEnKS)~\citep{bocquet2014iterative,sakov2012iterative}, IEnKS
with multiple data assimilation (IEnKS-MDA)~\citep{bocquet2013joint,bocquet2014iterative}, and
ensemble-based four-dimensional variational method (En4D-Var)~\citep{liu2008ensemble,gustafsson2014four}.
These ensemble-based DA methods represent uncertainty with a finite ensemble and assume an approximately multivariate-Gaussian state, allowing accurate inference even with relatively small ensemble sizes.
We refer the reader to \citet{carrassi2018data} for a review of common DA approaches. 
In the context of inertial cavitation, \citet{spratt2021characterizing} recently incorporated ensemble-based DA into IMR to reduce the number of forward simulations required for accurate characterization.
Specifically, the large simulation budgets associated with brute-force curve fitting are replaced by a modest ensemble, yielding a scalable rheometry approach.
Moreover, the hybrid En4D-Var formulation can provide robust property estimates with fewer repeated measurements per dataset when characterizing hydrogel mechanics~\citep{spratt2021characterizing,mancia2021acoustic,chu2025bayesian}.

\begin{figure}[h]
    \centering
    \includegraphics[scale=1]{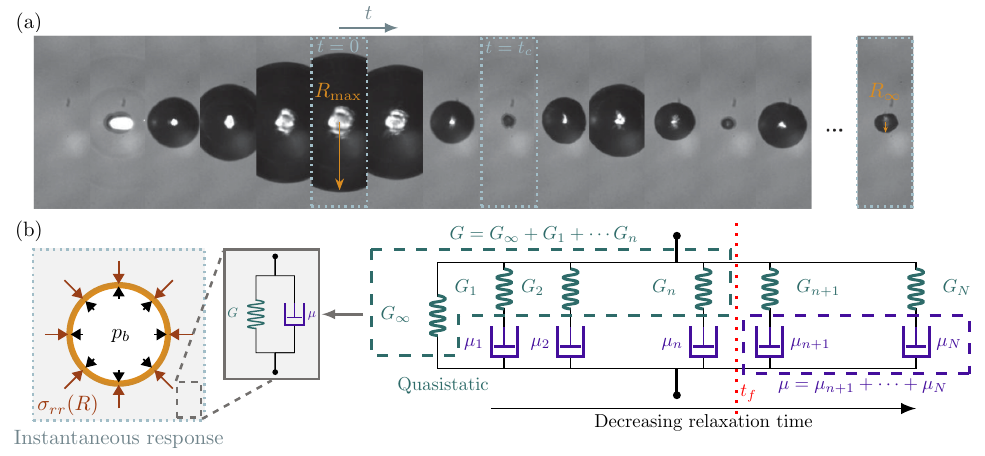}
    \caption{
        Schematic of bubble evolution during an LIC event: (a) experimental bubble snapshots; and (b) the stress response modeled by a lumped neo-Hookean Kelvin--Voigt model truncated from a multiple-branch Maxwell model.
        A single green laser pulse is reshaped and focused into a soft material sample.
        At the same time, an ultra-high-speed camera records the event at \SI{1e6}{\fps} to extract the radial dynamics, $R(t)$.
    }
    \label{f:evolution}
\end{figure}

A common assumption underlying these IMR-based techniques is that the general material response during cavitation, which may involve a spectrum of relaxation times, can be approximated by a frozen relaxation timescale, $t_f$, over the selected time horizon, thereby yielding a model with constant constitutive parameters.
\Cref{f:evolution} schematically illustrates this truncation from a multiple-branch, generalized Maxwell model to a lumped Kelvin--Voigt model with a single effective viscosity.
In practice, however, this assumption may be overly restrictive, because the medium surrounding the cavitation bubble may exhibit time- or state-dependent constitutive behavior, including thixotropic and rheopectic effects~\citep{cheng1965phenomenological, lee2009thixotropic}.
In LIC bubble dynamics, such behavior may be further amplified by the substantial variation in strain rate between collapse and subsequent rebounds, particularly around maxima.
In finite deformation viscoelasticity, the history- or state-dependence is commonly addressed by evaluating the stress through a hereditary integral~\cite{pipkin-rogers_1968,haupt-lion_2002} or by incorporating a set of evolving internal variables in the constitutive law~\cite{reese-govindjee_1998,bergstrom1998constitutive}.
In the LIC setting, IMR has been coupled with DA methods using short assimilation windows of up to a three-time-step lag to characterize transient constitutive behavior in PAAm gels throughout the cavitation process~\citep{spratt2021characterizing,buyukozturk2022particle}.
The trade-off associated with such short time horizons is increased uncertainty, as many parameter combinations can fit the observations equally well.
Deterministic, reduced-order, and data-assimilation variants of IMR all infer a single effective set of constant constitutive parameters.
We instead use time-resolved parameters to identify the stages of cavitation at which a constant-parameter model no longer reproduces the bubble dynamics.
Here, we introduce the modified IEnKS-MDA (MIEnKS-MDA) to enable accurate and robust inference of effective locally time-averaged material responses using substantially larger, highly overlapping assimilation windows.
After assimilation within each window, the full bubble state predicted by the IMR simulations is used to initialize the subsequent window.
Repetition of this posterior-to-forecast update over successive assimilation windows across the full time interval yields an estimate of the evolving constitutive properties.
Using the neo-Hookean Kelvin--Voigt (NHKV) model as an example, we show that MIEnKS-MDA enables a more faithful constitutive relation by improving agreement between measurements and simulations across different stages of bubble dynamics, including collapse and rebound.


In \cref{S:IMR}, we introduce the theoretical bubble dynamics model and the numerical methods used for forward simulation.
\Cref{S:IMR_methods} then describes standard IMR, which achieves material characterization by combining these simulations with LIC measurements.
We review existing reduced-order and DA variants of IMR and introduce IMR-MIEnKS-MDA, an IMR-based sliding-window DA technique for tracking the evolution of material properties throughout the cavitation process.
\Cref{S:results} reports the characterization results for two PAAm hydrogel formulations at different crosslinker concentrations and for a gelatin gel, each tested across three representative temperatures.
\Cref{s:limits,s:conclusions} summarize the main contributions and limitations.

\section{Inertial microcavitation rheometry for laser-induced cavitation}\label{S:IMR}

\begin{figure}[t]
    \centering
    \includegraphics[scale=1]{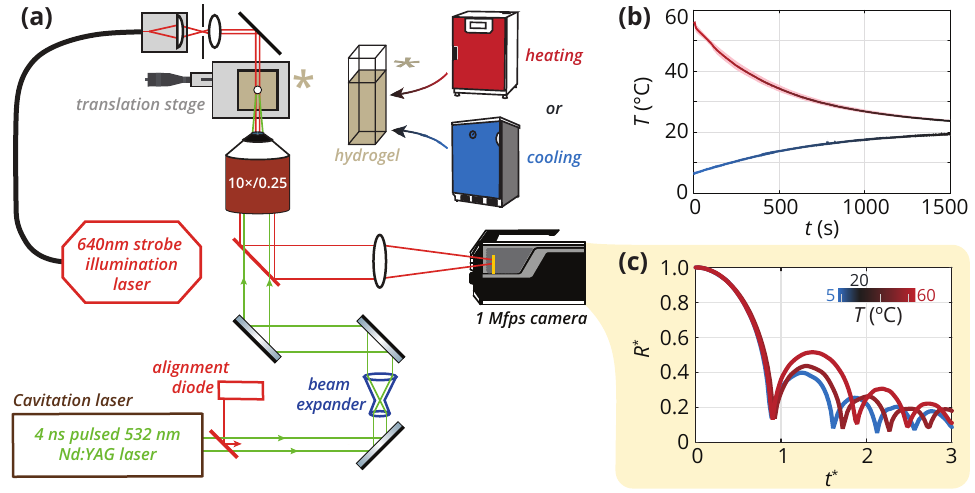}
    \caption{
        Schematic of the laser-induced cavitation (LIC) experiments:
        (a) a single green laser pulse is reshaped and focused into a hydrogel sample to nucleate an inertial microcavitation;
        (b) heated and cooled specimens undergoing free convection are monitored using thermocouples, enabling measurements over a wide temperature range; and
       (c) the radial bubble dynamics, $R(t)$, are extracted from images recorded by an ultra-high-speed camera at $\sim$1 million frames per second. 
     }
    \label{f:LIC}
\end{figure}

Our objective is to accurately and efficiently characterize the viscoelastic properties of hydrogels in high-strain-rate, uncertainty-prone regimes.
This section presents the IMR method as a high-strain-rate rheometer, achieved by integrating laser-induced cavitation (LIC) experimental measurements with numerical simulations of physical bubble dynamics models~\citep{estrada2018high}.
\Cref{f:LIC} presents a schematic illustrating the IMR methodologies for the quantitative characterization of soft materials using bubble radius–time histories obtained from LIC experiments.

\subsection{Theoretical bubble dynamics model}

The dimensionless Keller--Miksis equation~\citep{keller1980bubble} is applied to govern the spherically symmetric motion of bubble dynamics in a viscoelastic material assumed to be nearly incompressible. 
\begin{align}
    \left(1-\frac{\dot{R}^*}{c^*}\right)R^*\Ddot{R}^*+\frac{3}{2}\left(1-\frac{\dot{R}^*}{3c^*}\right)\dot{R}^{*^2}=\left(1+\frac{\dot{R}^*}{c^*}+\frac{R^*}{c^*} \dv{}{t}\right)\left(p_b^*-\frac{1}{\mathrm{We} \, R^*} +S^*-1\right).
    \label{eqn:keller-miksis}
\end{align}
The nondimensionalization is performed using the characteristic velocity, $U_c=\sqrt{p_{\infty}/\rho}$, and the material stretch ratio, $\lambda = R/R_{\infty}$.
The details of dimensionless parameters are summarized in \cref{dimensionless_quantity}. 
The bubble contents are assumed to consist of two components: water vapor and gas considered to be non-condensible, characterized by gas constants $\mathcal{R}_v$ and $\mathcal{R}_g$, on the time scales of inertial cavitation~\citep{akhatov2001collapse,nigmatulin1981dynamics}.
This mixture is assumed to be homobaric and follows the ideal gas law.
We assume that the mass and heat transfer of the gases within the bubble obeys Fick's law and Fourier's law.
The cavitation model begins when the bubble reaches its maximum radius and thermodynamic equilibrium, $R^*(0)=1$, to avoid the initial bubble growth phase.

\begin{table}[ht!]
    \centering
    \caption{Dimensionless quantities used in this manuscript.}\label{dimensionless_quantity}
    {\setlength{\tabcolsep}{9pt}
    \begin{tabular}{r l l}
        Dimensional & Dimensionless quantity  & Quantity name       \\ \midrule
         &  $\lambda = R/R_{\infty}$ & Material stretch ratio   \\
        $t$  &  $t^*=t U_c/R_\mathrm{max}$     & Time  \\
        $R$  &  $R^*=R/R_\mathrm{max}$ & Bubble-wall radius   \\
        $U=\dot{R}$  &  $U^*=U/U_c$ & Bubble-wall velocity   \\
        $R_{\infty}$  &  $R_{\infty}^*=R_{\infty}/R_\mathrm{max}$ & Equilibrium bubble-wall radius   \\
        $c$  &  $c^*=c/U_c$ & Material wave speed   \\
        $p_b$  &  $p_b^*=p_b/p_{\infty}$ & Bubble-wall pressure   \\
        $p_{\mathrm{v,\,sat}}(T_{\infty})$  & 
        $p_{\mathrm{v,\,sat}}^* = p_{\mathrm{v,\,sat}}(T_{\infty})/p_{\infty}$ & Vapor saturation pressure \\ 
        $C$  &  $C^*=\left[1+(\mathcal{R}_v/\mathcal{R}_g)(p_b^*/p^*_{\mathrm{v,\,sat}}-1)\right]^{-1}$ &  Vapor concentration   \\
        $T$  &  $T^*=T/T_{\infty}$ &  Temperature   \\
        $R_\mathrm{max}$  &  $\mathrm{We}=p_{\infty}R_\mathrm{max}/(2\gamma)$ & Weber number   \\
        $S$  &  $S^*=S/p_{\infty}$ & Stress integral   \\
        $G$  &  $1/\mathrm{Ca}=G/p_{\infty}$ & 1/Cauchy number   \\
        $\mu$  &  $1/\mathrm{Re}=\mu/(\rho U_c R_{\mathrm{max}})$ & 1/Reynolds number   
    \end{tabular}
    }
\end{table}

The instantaneous forcing on the bubble wall is governed by the bubble-wall pressure, $p_b^*$, the viscoelastic stress integral of the surrounding material, $S^*$, and surface-tension contributions, as illustrated schematically in \cref{f:LIC}.
The pressure inside the bubble is coupled to the internal energy equation~\citep{barajas2017effects,estrada2018high}, while the surface-tension effect depends on the bubble-wall velocity, $\dot{R}^*$.
The Kelvin--Voigt (KV) model captures linear viscoelastic behavior, and its neo-Hookean Kelvin--Voigt (NHKV) extension incorporates a neo-Hookean elastic term, enabling the model to represent nonlinear viscoelastic responses~\citep{gaudron2015bubble}.
The NHKV model can be interpreted as a Taylor-series approximation of the more general Fung model~\citep{fung2013biomechanics}, and is therefore well-suited for describing nonlinear viscoelastic behavior at high strain rates~\citep{yang2022mechanical}.
The stress integral associated with the NHKV model takes the form of
\begin{align}
    S^*_{\mathrm{NHKV}} &= {-\frac{4 U^*}{\mathrm{Re} R^*}-\frac{1}{2\mathrm{Ca}}
    \left[5-\frac{4}{\lambda}-\frac{1}{\lambda^4}\right]} \label{eqn:stress integral_NHKV}.
\end{align}
For more details, readers are referred to \citet{estrada2018high}.
In this work, we focus on the rheometry based on the NHKV model.
In the following, all variables are nondimensionalized according to \cref{dimensionless_quantity}, and we omit the superscript $(\cdot)^*$ for brevity.

\subsection{Numerical methods}

Using the governing equations together with the prescribed maximum and equilibrium bubble radii, we numerically integrate the bubble dynamics system forward in time.
The state vector is
\begin{align}\label{eq:state}
    \vb*{q}(t) = \left\{R,\,\dot{R},\, p_b,\,S,\,T,\,C,\,\{\mathcal{M},\vb*{\phi}_{\mathcal{M}}\}\right\},
\end{align}
where the state parameters represent the bubble-wall radius, velocity, bubble pressure, stress integral, the discretized temperature and vapor concentration fields inside the bubble, the candidate constitutive model, $\mathcal{M}$, with its associated material-parameter set, $\vb*{\phi}_{\mathcal{M}}$.
We further introduce a modeling parameter 
\begin{align}
    \vb*{\theta} \equiv \{\mathcal{M},\vb*{\phi}_{\mathcal{M}}\}
\end{align}
to denote the combination of model choice and its corresponding material properties.
Following \citet{estrada2018high}, we treat the far-field density, pressure, and temperature as constants for the physical regime of interest, though this assumption is not a restriction of the method.

The discrete-time nonlinear dynamical system can be written as 
\begin{subequations}\label{IMR_det}
\begin{align}
    \vb*{q}_{k+1} &= \mF_k(\vb*{q}_k; \vb*{\theta}), \\
     R_{k+1} &= \mH(\vb*{q}_{k+1}),
\end{align}
\end{subequations}
where $\mF_k$ is the nonlinear evolution operator for the full bubble state, parameterized by the constitutive parameters $\vb*{\theta}$.
A linear observation function $\mH$ is used to map the state $\vb*{q}$ to a point in the measurement space. 
In this study, we take the bubble radius $R^*$ as the primary observable variable because it is directly measurable in experiments.
Over a time horizon $t\in\left[ 0, t_{N_t} \right]$ discretized into $N_t$ time steps, we collect the deterministic state trajectory and corresponding observations as $\vb*{Q}_{0\to t_{N_t}} = \mqty[ \vb*{q}_0 &  \vb*{q}_1 & \cdots & \vb*{q}_{N_t}]$ and $\vb*{Y}_{0\to t_{N_t}} = \mqty[ R_0 & R_1 &\cdots & R_{N_t}]$, respectively, yielding
\begin{align}\label{IMR_det_all}
    \vb*{Q}_{0\to t_{N_t}} = \mF(\vb*{\theta}) 
    \quad \text{and} \quad
    \vb*{Y}_{0\to t_{N_t}} = \mH(\vb*{Q}_{0\to t_{N_t}}), 
\end{align}
where $\mF$ represents the nonlinear operator that creates the space-time state sequence over $t \in \left[0, t_{N_t} \right]$.
In practice, we solve the resulting stiff ODE system using the \texttt{MATLAB} solver \texttt{ode23t}, and the solution is sampled at a time interval of $\Delta t^*=0.025$ for consistency with the data-assimilation procedure.
We then use this deterministic forward model to identify the optimal constitutive parameters within the parameter space, $\vb*{\theta}\in \Theta$, by minimizing the mismatch between the model predictions and the LIC measurements.
The experimental measurements, originally sampled uniformly in physical time, are interpolated 
onto the same uniform nondimensional time grid, denoted by $\tilde{(\cdot)}$.

\section{IMR-based characterization techniques}\label{S:IMR_methods} 

We seek accurate and efficient characterization approaches that are also robust to the uncertainty inherent in the $N_\mathrm{exp}$ repeated measurements.
In this section, we introduce several variants of IMR for calibrating material properties, including the standard, reduced-order (parsimonious), and data assimilation (DA) versions.
As these variants assume material properties remain invariant over the cavitation event, we further propose a sliding-window rheometer that can accommodate apparent changes in material behavior.
An overview of these methods is provided in \cref{IMR_methods}.
We note that the sliding-window method incurs the highest simulation cost of the techniques considered; this cost is paid for time resolution rather than speed, arising from the repeated ensemble propagation across the $N_\mathrm{w}$ assimilation windows.
Unlike brute-force IMR, however, the expense does not grow with the resolution of the parameter space, and the ensemble propagation within each window is embarrassingly parallel.

\begin{table}[ht!]
    \centering
    \caption{
        Overview of IMR-based characterization techniques, including their descriptions, inference-window sizes, uncertainty treatment as deterministic (D) or statistical (S), computational cost for a 2-parameter model per LIC measurement in terms of simulation steps, and constitutive properties throughout the cavitation.
    }
    \begin{tabular}{l c c c c c c}\label{IMR_methods}
        \multirow{2}{*}{Rheometry} & \multirow{2}{*}{Description}  & Inference   & \multirow{2}{*}{D/S}  & Simulation & Model & \multirow{2}{*}{\S}  \\ 
        & &window &  & Steps & parameters & \\\midrule
        IMR~\citep{estrada2018high}  & standard & $\left[0, t_{N_t}\right]$ & D & $\sim O(10^6)$ & constant  & \ref{sec:IMR}\\
        pIMR~\citep{zhu2025parsimonious} & reduced-order & $t=t_c$ & D  & No Sims & constant & \ref{sec:pIMR}\\
        En4D-Var~\citep{spratt2021characterizing} & single-window DA & $\left[0, t_{N_t}\right]$ & S  & $\sim O(10^5)$ & constant & \ref{sec:En4Dvar} \\
        MIEnKS-MDA  & sliding-window DA & $\left[t_{jh}, t_{jh+L-1}\right]$ & S & $\sim O(10^7)$ & evolving & \ref{sec:IEnKS}\\
    \end{tabular}
\end{table}

\subsection{Deterministic approaches}

\subsubsection{Standard IMR}\label{sec:IMR}

The standard IMR seeks to minimize the discrepancy between simulated and experimentally measured bubble dynamics over a prescribed time window, typically chosen from the equilibrium state through the second or third bubble collapse.
The parameters are obtained by solving the optimization problem that minimizes a cost function, defined as
\begin{align}
    J_{\mathrm{IMR}}(\vb*{\theta}) =
    \frac{1}{N_\mathrm{exp}} \sum_{n=1}^{N_\mathrm{exp}}  {\left\lVert \vb*{Y}_{0\to t_{N_t}}(\vb*{\theta})-\btY_{0\to  t_{N_t}}^{(n)}\right\rVert}^2.
\end{align}
By construction, standard IMR yields an effective, time-averaged characterization over the selected window by identifying the optimal parameter $\vb*{\theta}^\star$, associated with the minimum cost, $J^\star$.
However, the strong nonlinearity of the bubble dynamics generally necessitates dedicated search algorithms to identify the optimum, making brute-force parameter estimation computationally expensive,
particularly when processing large datasets while accounting for measurement uncertainty.
In practice, robust identification often requires repeated measurements of $N_\mathrm{exp}=O(10)$~\citep{estrada2018high,yang2020extracting,spratt2021characterizing}.

\subsubsection{Parsimonious IMR}\label{sec:pIMR}

Parsimonious IMR~(pIMR) offers a computationally efficient alternative for material characterization by changing the inference target from the full bubble-radius trajectory to a single scalar observable, the bubble collapse time~\citep{zhu2025parsimonious}.
We rewrite the non-dimensional Keller--Miksis equation \cref{eqn:keller-miksis} in a compact form,
\begin{equation}
    R \ddot{R} + \frac{3}{2} \dot{R}^{2} = -1 +  {f}\left( t, R, \dot{R}, \ddot{R} ; \vb*{\theta}\right).
    \label{eqn:approx-keller-miksis}
\end{equation}
Here, the modifier $f$ acts as an effective pressure term that collects non-inertial contributions, including viscoelastic stresses, surface tension, and other constitutive or forcing effects, thereby coupling these physical mechanisms to the bubble dynamics.
Following the analysis of \citet{zhu2025parsimonious}, we approximate the collapse time without performing forward simulations as
\begin{equation}
    {t}_c(\vb*{\theta}) = 0.9147 \left( 1 - \overline{f}(\vb*{\theta}) \right)^{-1/2},
    \label{eqn:collapse-time-approx}
\end{equation}
where $\overline{f}$ is the temporal average of $f$ during the collapse process.
The prefactor of $0.9147$ is the Rayleigh collapse time for a void in an inviscid medium driven solely by the far-field pressure.
Details of the analysis procedure and the resulting functional forms of $\overline{f}$ for various physical mechanisms are summarized in \ref{App:pIMR}.
Furthermore, a more thorough discussion is available in \citet{zhu2025parsimonious}.

Specifically, pIMR minimizes the mismatch between the time from maximum bubble size to its first minimum---termed the first collapse time---predicted by \cref{eqn:collapse-time-approx} and the experimentally measured value, $\tilde{(\cdot)}^{(n)}$.
This optimization is achieved by minimizing the pIMR cost function 
\begin{align}
    J_{\mathrm{pIMR}}(\vb*{\theta}) =
    \frac{1}{N_\mathrm{exp}} \sum_{n=1}^{N_\mathrm{exp}} 
    \left[ \left(\frac{\tilde{t}_c^{{(n)}}}{{t}_c(\vb*{\theta})} \right)^2 - 1 \right]^2. 
\end{align}
In practice, we minimize the logarithm of $J_{\mathrm{pIMR}}$ to improve numerical conditioning as $J_{\mathrm{pIMR}}$ can vary over several orders of magnitude across the parameter space.
Compared to the standard IMR approach introduced in \cref{sec:IMR}, evaluating $J_{\mathrm{pIMR}}$ requires no forward simulations of the bubble dynamics, enabling rapid exploration of the parameter space $\Theta$.
Compared with IMR, which uses full bubble dynamics within the first-collapse window, relying on collapse time alone can increase inferential uncertainty, particularly in the presence of measurement noise.
Experimentally, variability in the measured collapse times may complicate the identification of a single best-fit value, motivating statistical formulations that explicitly account for these uncertainties.


\subsection{Statistical approaches: ensemble-based data assimilation for IMR}

Beyond the deterministic approaches described above, data assimilation (DA) provides a statistical framework for systematically accounting for uncertainty and measurement noise by inferring probabilistic parameter distributions.
Specifically, we assume the relevant variables, such as shear modulus and viscosity, follow multivariate Gaussian distributions and model them using an ensemble of realizations.

The ensemble-based DA approaches consist of a forecast step followed by an analysis step.
In the forecast step, we draw an ensemble of $N_{\mathrm{En}}$ parameter instances from the prior distribution,
\begin{align}
    \vb*{\theta}\sim  \mathcal{N}(\vb*{\mu}_{\theta}^{(0)},\vb*{\Sigma}_{\theta}^{(0)}),
\end{align}
and construct an ensemble of $N_{\mathrm{En}}$ realizations at time step $l$,
\begin{align}
    \vb*{Q}_l (\vb*{\Theta}_l) = \mqty[ \vb*{Q}_l\left(\vb*{\theta}^{(1)} \right) & \cdots & \vb*{Q}_{l}\left(\vb*{\theta}^{(N_\mathrm{En})}\right)].
\end{align}
Denoting the ensemble average by $\left<\cdot\right>$, we define the scaled state perturbation matrix
\begin{align}
    \vb*{Q}'_l = \frac{1}{\sqrt{N_{\mathrm{En}}-1}}
    \mqty[ \vb*{Q}_l^{(1)}-\left<\vb*{Q}_l\right>  & \cdots & \vb*{Q}_{l}^{(N_{\mathrm{En}})}-\left<\vb*{Q}_l\right>],
\end{align}
and the corresponding ensemble covariance matrix $\vb*{C}_l=\vb*{Q}'_l {\vb*{Q}'_l}^\top$, which characterizes the multivariate Gaussian distribution implied by the ensemble.
We propagate each ensemble member forward in time using the IMR solver \cref{IMR_det}.
The corresponding ensemble predictions of the bubble radius at time step $k$ are collected as
 \begin{align}
     \vb*{Y}_k(\vb*{Q}_l) = \mqty[ {R_k}^{(1)} & \cdots &  {R_k}^{(N_\mathrm{En})}].
 \end{align}
 We consider the weighted norms for the observation and state spaces,
\begin{align}
    \|\vb*{Y}_k\|^2_{\vb*{P}_k} \equiv \vb*{Y}_k^\top \vb*{P}_k^{-1}\vb*{Y}_k
    \quad \text{and} \quad  
    \|\vb*{Q}_l\|^2_{\vb*{C}_l} \equiv \vb*{Q}_l^\top \vb*{C}_l^{-1}\vb*{Q}_l ,
\end{align}
where $\vb*{P}_k$ is the measurement noise covariance matrix at time step $k$.
The analysis step then quantifies the discrepancy between these predictions and the experimental measurements, updating the ensemble accordingly.

\subsubsection{IMR-based En4D-Var}\label{sec:En4Dvar}

The ensemble-based four-dimensional variational (En4D-Var) approach has demonstrated its computational efficiency by iteratively updating the initial ensemble at time step $l=0$~\citep{spratt2021characterizing,mancia2021acoustic,chu2025bayesian}.

For each measurement, the associated En4D-Var cost function is defined as
\begin{align}\label{cost_en4dvar}
    J_{0\to  t_{N_t}}^{(n)}(\vb*{Q}_0(\vb*{\Theta}_0)) = \frac{1}{2 N_t} \sum_{k=1}^{N_t} \left\lVert\btY_k^{(n)}-\vb*{Y}_k(\vb*{Q}_0)\right\rVert^2_{\vb*{P}_k} 
    + \frac{1}{2}\left\lVert\vb*{Q}_0-\left<\vb*{Q}_0\right>\right\rVert^2_{\vb*{C}_0},
\end{align}
which comprises a data-misfit term and a regularization term that penalizes deviations of the initial ensemble from its mean, weighted by the prior covariance.
Details of the optimization of \cref{cost_en4dvar} are provided in \ref{App:En4dVar}.
As $\vb*{Q}_0$ is updated iteratively through En4D-Var, the associated ensemble of modeling parameters $\vb*{\theta}$ is updated accordingly. 
For each measurement, we approximate the resulting parameter ensemble by a multivariate Gaussian,
\begin{align}\label{eqn:Gaussain_exp_n}
    \vb*{\theta}^{(n)}\sim  \mathcal{N}(\vb*{\mu}_{\theta}^{(n)},\vb*{\Sigma}_{\theta}^{(n)}).
\end{align}
Assuming independent measurements, the fused posterior is therefore also Gaussian and can be expressed as
\begin{align}
    \vb*{\theta}\sim  \mathcal{N}(\vb*{\mu}_{\theta},\vb*{\Sigma}_{\theta}),
\end{align}
where its mean and variance are computed as
\begin{align}\label{eqn:posterier_en4d}
    \vb*{\Sigma}_{\theta}^{-1} &\equiv \sum_{n=1}^{N_\mathrm{exp}} \left(\vb*{\Sigma}^{(n)}_{\theta}\right)^{-1}-(N_\mathrm{exp}-1)\left({\vb*{\Sigma}^{(0)}_{\theta}}\right)^{-1}, \quad \text{and}\\
    \vb*{\mu}_{\theta} &\equiv \vb*{\Sigma}_{\theta}  \left[\sum_{n=1}^{N_\mathrm{exp}}\left(\vb*{\Sigma}^{(n)}_{\theta}\right)^{-1}\vb*{\mu}_{\theta}^{(n)} -(N_\mathrm{exp}-1){\vb*{\Sigma}_{\theta}^{(0)}}^{-1}\vb*{\mu}^{(0)}_{\theta}\right],
\end{align}
respectively. 
The optimal parameter estimate for En4D-Var is then defined as the posterior mean, $\vb*{\theta}^\star=\vb*{\mu}_{\theta}$.
In practice, we use an ensemble size of $N_\mathrm{En}=48$, which has been shown to provide accurate material-property estimates~\citep{spratt2021characterizing,mancia2021acoustic}, and couple the assimilation with a restart strategy that further reduces computational cost and prior-induced bias~\citep{chu2025bayesian}.
The DA hyperparameters used throughout this work are summarized in \cref{tab:DA_params} of \ref{App:En4dVar}.

Compared to standard IMR, En4D-Var provides a statistical characterization by inferring parameter distributions from discrete ensembles, substantially reducing computational cost by avoiding brute-force exploration of the full parameter space.
By construction, both rheology approaches characterize time-averaged behavior over the prescribed time window. 
However, the observed dependence of the inferred properties on window length in both methods suggests that the material parameters we infer may \textit{themselves} depend on time; hence, our model choice/parameterization (implicitly assuming constant parameters) is insufficient to describe or predict material response. 
To address this limitation, we seek to infer the time evolution of constitutive properties and thereby improve both the description of material behavior during cavitation and the identification of the possible regimes in which we must refine our material model, as described next.

\subsubsection{IMR-based modified IEnKS-MDA}\label{sec:IEnKS}

\begin{figure}[h]
    \centering
    \includegraphics[scale=1]{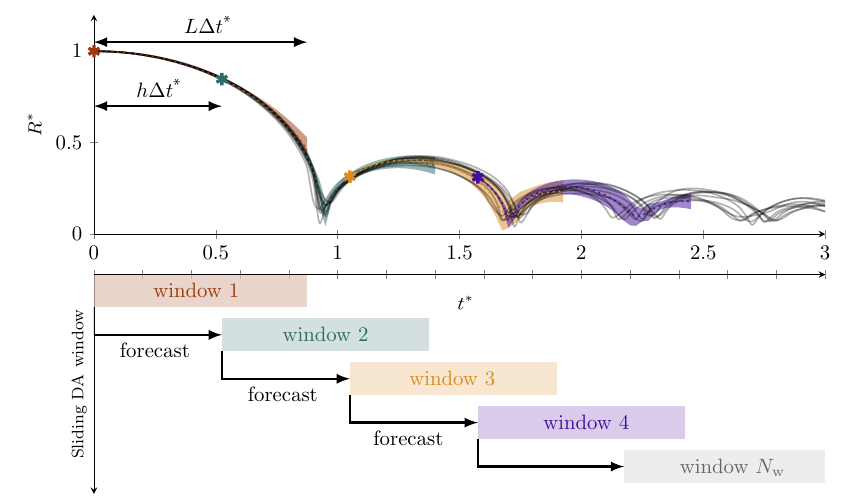}
    \caption{
        Schematic of the MIEnKS-MDA process for constitutive-property characterizations from LIC measurements.
        After assimilating the data within the current time window, the predicted full bubble state, including radius, pressure, and temperature, is used to initialize the subsequent assimilation window, separated by a time gap of $h\Delta t^*$. 
        This posterior-to-forecast update is repeated over the entire time interval, yielding an estimate of the evolving constitutive properties.
    }
    \label{f: sliding DA overview}
\end{figure}

Building on the iterative ensemble Kalman smoother with multiple data assimilation (IEnKS-MDA)~\citep{bocquet2013joint,bocquet2014iterative}, we introduce a sliding-window rheometer for tracking evolving constitutive properties.
We partition the full time interval $t\in\left[0, t_{N_t}\right]$ into successive, overlapping assimilation windows of length $L$ and shift $h$.
The $j$th window spans 
\begin{align}
    t\in\left[t_{(j-1)h}, t_{(j-1)h+L-1}\right],
\end{align}
yielding a total of $N_\mathrm{w}=\lfloor (N_t-L+1)/h\rfloor+1$ windows.
Analogous to the En4D-Var objective in \cref{cost_en4dvar}, we define the cost function over the $j$th window as 
\begin{align}\label{cost_q}
    J_{j}^{(n)}(\vb*{Q}_j(\vb*{\Theta}_j)) = \frac{1}{2 L} \sum_{k=(j-1)h}^{(j-1)h+L-1} \left\lVert\btY_k^{(n)}-\vb*{Y}_k(\vb*{Q}_j)\right\rVert^2_{\vb*{P}_k} 
    + \frac{1}{2}\left\lVert\vb*{Q}_j-\left<\vb*{Q}_j\right>\right\rVert^2_{\vb*{C}_j}.
\end{align}
The posterior from window $j$ is used to initialize window $j+1$.
Specifically, we propagate the $N_\mathrm{En}$ ensembles from $t_{(j-1)h}$ to overlap time $t_{jh}$ and use the resulting ensembles as the initial forecast condition for the next window.
Whereas standard IMR initializes the full bubble state $\vb*{q}$, defined in \cref{eq:state}, by assuming that the bubble has reached its maximum radius and thermodynamic equilibrium, the present sliding-window rheometer predicts the state directly from the numerical simulation and therefore retains the evolved stress, temperature, pressure, and other internal variables.
The subsequent window assimilates the states $\vb*{Q}_{j+1}$ and the material properties $\vb*{\Theta}_{j+1}$ using observations on $t\in\left[t_{jh}, t_{jh+L-1}\right].$
The overlap between successive windows, together with this posterior-to-forecast initialization, promotes continuity in the inferred material response.
This initialization is the modification referenced in the method name: whereas previous IMR-based implementations of IEnKS-MDA re-initialized each window from an assumed equilibrium state at the maximum bubble radius~\citep{spratt2021characterizing,buyukozturk2022particle}, the present scheme hands the evolved internal state off between long, overlapping windows.
A schematic of the modified IEnKS-MDA~(MIEnKS-MDA) process is shown in \cref{f: sliding DA overview}.

After completing the MIEnKS-MDA procedure, we fuse the posteriors from the $N_{\mathrm{exp}}$ independent measurements for each window $j$, analogously to \crefrange{eqn:Gaussain_exp_n}{eqn:posterier_en4d}.
This fusion yields a Gaussian posterior for the window-wise constitutive parameters,
\begin{align}
    \vb*{\theta}_j \mid \{\vb*{Y}^{(n)}\}_{n=1}^{N_{\mathrm{exp}}}
    \sim \mathcal{N}\!\left(\vb*{\mu}_{\theta,j},\,\vb*{\Sigma}_{\theta,j}\right).
\end{align}
To obtain time-resolved material properties, we combine the window-wise posteriors from all windows that contain a given time step $t_k$.
We define the index set
\begin{align}
    \mathcal{I}(k) \coloneqq \Bigl\{\, j \;:\; (j-1)h \leq k \leq (j-1)h+L-1 \,\Bigr\}.
\end{align}
As adjacent windows overlap, their posterior estimates are generally correlated.
We therefore use a normalized precision-weighted aggregation rather than treating the overlapping windows as independent sources of information.
The resulting local posterior summary is written as
\begin{align}
    \vb*{\theta}(t_k) \sim \mathcal{N}\!\left(\vb*{\mu}_{\theta}(t_k),\,\vb*{\Sigma}_{\theta}(t_k)\right),
\end{align}
with 
\begin{align}
    \vb*{\Sigma}_{\theta}(t_k)^{-1} &= \sum_{j\in\mathcal{I}(k)} w_{j,k}\vb*{\Sigma}_{\theta,j}^{-1}, \quad \text{and} \quad 
    \vb*{\mu}_{\theta}(t_k) = \vb*{\Sigma}_{\theta}(t_k)\sum_{j\in\mathcal{I}(k)} w_{j,k} \vb*{\Sigma}_{\theta,j}^{-1}\vb*{\mu}_{\theta,j}.
\end{align}
Here, the weights $w_{j,k}=1/|\mathcal{I}(k)|$ normalize the contribution of the windows containing $t_k$, thereby preventing the effective precision from being overcounted.
This aggregation provides a smoothed time-resolved estimate while retaining the relative confidence of the individual window-wise posteriors.

In practice, we set the window shift to $h=5$ to balance computational cost against continuity between successive assimilation windows.
With this choice, the total number of windows $N_{\mathrm{w}}$ is reduced by approximately a factor of five relative to the minimum shift $h=1$.
Previous IMR studies have coupled IEnKS-MDA with short-lag windows (up to $L=3$) to resolve transient behavior~\citep{spratt2021characterizing,buyukozturk2022particle}.
Such limited horizons may lead to substantial uncertainty, as many parameter sets can reproduce the observations.
Here, we instead use a substantially larger assimilation window size of $L=35$ to propagate the full internal state between windows and thereby estimate effective, time-averaged responses, while keeping $L \Delta t^*=0.875$ comparable to the first-collapse time to mitigate over-fitting.
A convergence study for inferred material properties is provided in \cref{f:PA property evolution_convergence}.

\begin{figure}[ht!]
    \centering
    \includegraphics[scale=1]{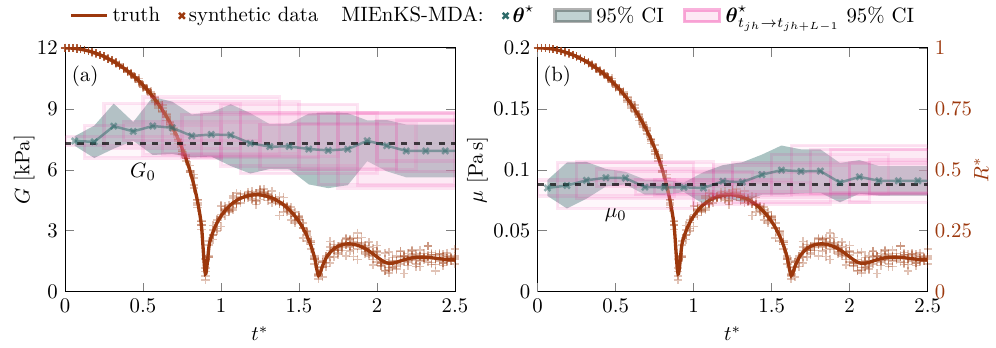}
    \caption{
        Constitutive properties inferred from $5$ sets of synthetic data about the prescribed ground truth, simulated using NHKV parameters $G_0$ and $\mu_0$.
       The MIEnKS-MDA estimates represent locally averaged transient material responses, with the properties within each sliding assimilation window also highlighted.
    }
    \label{f: verification}
\end{figure}

To verify the MIEnKS-MDA approach for rheometry, we generate synthetic data about a prescribed ground truth obtained from an IMR simulation with bubble parameters $R_\mathrm{max}=\SI{263}{\micro\meter}$ and $R_\mathrm{max}/R_{\infty}=7.58$, and constant NHKV model parameters of
$G_0=\SI{7.29}{\kilo \pascal}$ and $\mu_0 =\SI{0.088}{\pascal \second}$.
Specifically, time- and state-dependent synthetic noise is added to the simulated bubble dynamics, with standard deviations of $\sigma = |R^*-1|/20+(1-\exp( -t^*/10))/30$, to emulate experimental uncertainty~\citep{CHU2026115002}.
\Cref{f: verification} shows the resulting estimated constitutive properties throughout the synthetic cavitation processes.
The MIEnKS-MDA approach provides accurate estimates close to the prescribed NHKV ground truth, with only weak dependence on the cavitation process.
These results verify the ability of the proposed method to recover the underlying constitutive parameters from noisy, transient bubble-dynamics observations.
Moreover, because the synthetic data are generated with constant material parameters, this verification case suggests that any strong cavitation-dependent variation inferred from experimental data is unlikely to arise solely from sliding-window inference.
Instead, it may reflect physical changes of the material response during cavitation.


\section{Results} \label{S:results}


To demonstrate the characterization results, we examine three hydrogel formulations: stiff PAAm ($5\%$w/w acrylamide with $0.3\%$w/w bis-acrylamide), soft PAAm ($5\%$w/w acrylamide with $0.03\%$w/w bis-acrylamide), and gelatin ($5\%$w/w).
Fabrication protocols are provided in \ref{App:Materials}.
These materials were selected to provide different thermal responses: PAAm is expected to exhibit low temperature sensitivity over the range studied, whereas gelatin is expected to undergo a phase transition near $\SI{30}{\degreeCelsius}$.


\begin{table}[ht!]
    \centering
    \caption{
    Summary of the material samples used in the LIC experiments and the corresponding measured quantities, including temperature, bubble radius, and first-collapse time.
    }\label{datasets}
    \begin{tabular}{l c c c c c c c}
        Material   & Concentration   & $T$~[\SI{}{\degreeCelsius}] & $N_\mathrm{exp}$  & $R_\mathrm{max}~[\SI{}{\micro\meter}]$ & $R_\mathrm{max}/R_{\infty}$ & $t_c^*$ & \S \\ \midrule
        \multirow{6}{*}{PAAm}  & \multirow{3}{*}{\shortstack[c]{Acry/Bis \\ $5\%/0.3\%$}} & $10.1\pm 1.0$ & $30$ & $386\pm124$ & $8.08\pm1.06$ &$0.861\pm0.023$ & \multirow{3}{*}{\ref{sec:PA0503}} \\ 
         &   &  $\sim 21$ & $57$ & $318\pm162$  & $7.93\pm1.09$  &$0.882\pm0.029$ & \\ 
         &   &  $33.1\pm 1.7$ & $30$ & $371\pm129$ & $8.07\pm1.04$ &$0.891\pm0.020$ & \\ \cmidrule(lr){2-8}
         &\multirow{3}{*}{\shortstack[c]{Acry/Bis \\ $5\%/0.03\%$}}  & $10.1\pm1.1$  &$30$ & $245\pm121$ & $8.35\pm 0.65$ & $0.931\pm0.034$ &  \multirow{3}{*}{\ref{sec:PA05003}} \\ 
         &   &  $\sim 21$ & $30$ & $266\pm153$  & $8.32\pm 0.61$ &$0.927\pm0.037$ & \\ 
         &   &  $30.0\pm 2.0$ & $20$ & $295\pm 58$ & $8.36\pm 0.51$ &$0.935\pm0.023$ & \\ \midrule
        \multirow{3}{*}{Gelatin}   & \multirow{3}{*}{$5\%$} &  $14.7\pm1.1$ & $18$  & $277\pm48$ & $7.09\pm 0.36$
        & $0.890\pm0.044$ & \multirow{3}{*}{\ref{sec:gelatin}}\\
        &  & $24.4\pm1.3$ & $14$ & $298\pm 59$ & $7.37\pm 1.28$ &$0.968\pm0.036$ & \\
        & & $32.8\pm 1.4$ & $7$ & $312\pm 32$ & $6.83\pm 0.49$
        & $0.982\pm0.038$ & \\
    \end{tabular}
\end{table}

All the specimens were tested under three thermal regimes: steady state at room temperature~($\sim \SI{21}{\degreeCelsius}$), cooling to room temperature after equilibration in a $\SI{60}{\degreeCelsius}$ water bath, and warming to room temperature after equilibration in a refrigerator at approximately $5\unit{\degreeCelsius}$.
All the microcavitation events were nucleated and imaged \citep{abeid2024experimental}, and post-processed \citep{estrada2018high} using previous methods.
For transient temperature conditions, inertial microcavitation was conducted over the span of 
$15$--$20$~\SI{}{\minute}, allowing IMR measurements to be collected over a temperature span of approximately $\SI{50}{\degreeCelsius}$.
The datasets for each material were binned using a temperature tolerance of $\pm\SI{2}{\degreeCelsius}$.
Under these protocols, bins below room temperature are reached on the warming branch, whereas bins above room temperature, including the gelatin bins at $\SI{24.4}{}$ and $\SI{32.8}{\degreeCelsius}$, are reached on the cooling branch following equilibration at $\SI{60}{\degreeCelsius}$.
In the analysis below, we focus on data collected at three representative temperatures per formulation, and \cref{datasets} summarizes the resulting datasets.

\begin{figure}[ht!]
    \centering
    \includegraphics[scale=1]{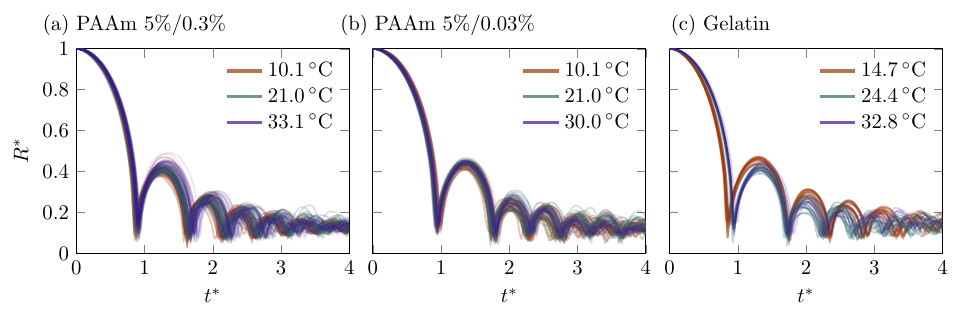}
    \caption{
        Bubble dynamics trajectories for the LIC process within three hydrogel formulations: (a) stiff PAAm; (b) soft PAAm; (c) gelatin $5\%$.
    }
    \label{f:Rt}
\end{figure}

\Cref{f:Rt} shows the bubble dynamics in the three considered gels.
Even before performing material characterization, the kinematic response of PAAm to LIC indicates limited temperature sensitivity over the tested range.
By contrast, LIC-induced kinematics of gelatin exhibit a clear temperature dependence: the dynamics at $\SI{24.4}{\degreeCelsius}$ and $\SI{32.8}{\degreeCelsius}$ are similar to one another, but differ visibly from those at $\SI{14.7}{\degreeCelsius}$.
In the following, we combine the measured bubble dynamics with the IMR-based characterization techniques introduced in \cref{S:IMR} to infer the material response during cavitation.
Specifically, the standard IMR fitting is performed over the parameter domain $G,\mu \in[0,25]~\SI{}{\kilo \pascal}\times [0,0.5]~\SI{}{\pascal \second}$, discretized with uniform resolutions of $\Delta G\approx\SI{0.21}{\kilo \pascal}$ and $\Delta \mu\approx\SI{0.0068}{\pascal \second}$, respectively.
The DA procedures are initialized from an uncorrelated multivariate normal prior with means of $\SI{6}{\kilo \pascal}$ and $\SI{0.209}{\pascal \second}$ and standard deviations of $\SI{2.8}{\kilo \pascal}$ and $\SI{0.088}{\pascal \second}$ for $G$ and $\mu$, respectively; note that this prior differs from the ground-truth values used in the synthetic verification of \cref{f: verification}.

\subsection{Polyacrylamide (PAAm) gels}\label{sec:PA0503}

\begin{figure}[ht!]
    \centering
    \includegraphics[scale=1]{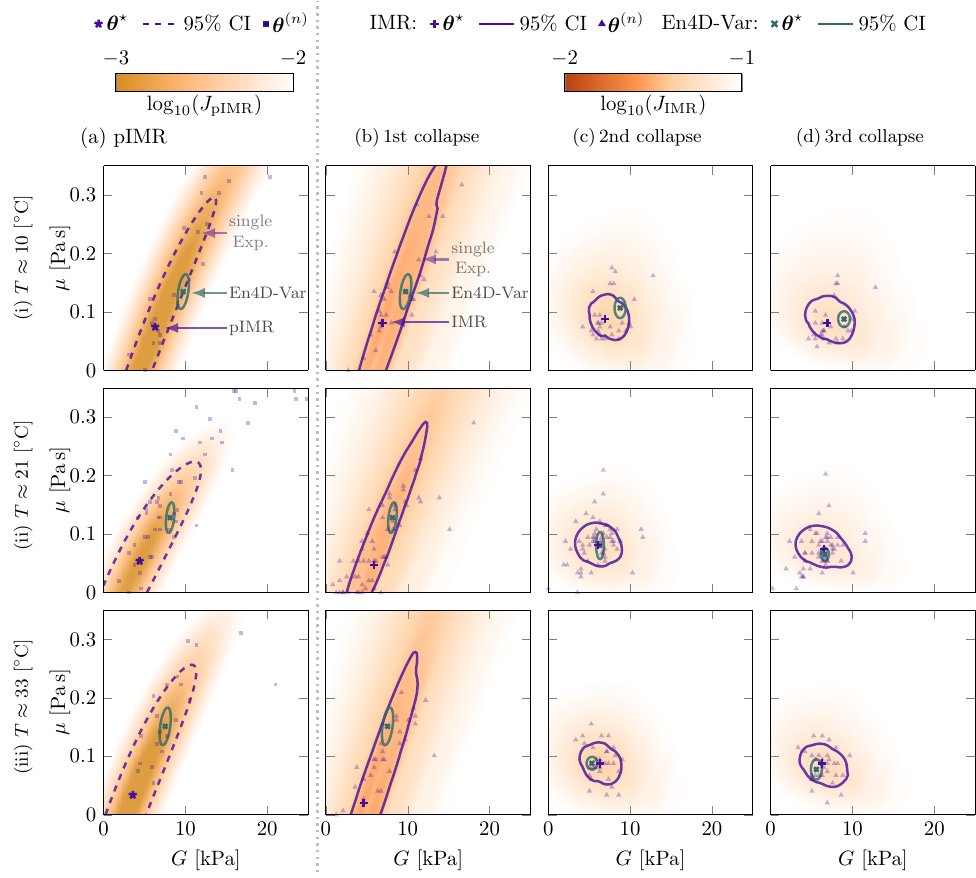}
    \caption{
        Calibrated material properties of stiff PAAm gels at different temperatures, (i) $T\approx$~\SI{10}{\degreeCelsius}; (ii) $T\approx$~\SI{21}{\degreeCelsius}; and (iii) $T\approx$~\SI{33}{\degreeCelsius}.
        Panel (a) shows the pIMR results, while panels (b)–(d) show the standard IMR and IMR-En4D-Var results obtained with fitting windows extending to the 1st, 2nd, and 3rd bubble collapses, respectively.
        Background contours show the cost-function values in the material-parameter space for pIMR and IMR, along with the corresponding single-experiment estimates, $\vb*{\theta}^{(i)}$, the overall optimal parameters, $\vb*{\theta}^\star$, and the associated 95\% confidence contours.
        The mean and 95\% confidence region for the inferred multivariate Gaussian distributions are shown for En4D-Var.
    }
    \label{f:PA0503_contour}
\end{figure}

As an illustrative example, \cref{f:PA0503_contour} compares the material calibration landscapes of stiff PAAm gels characterized using pIMR, IMR, and IMR-En4D-Var.
The relatively small variations in the identified material properties with temperature are consistent with the expected low-temperature sensitivity of PAAm gels.
Qualitatively, pIMR and the IMR results obtained using data up to the first bubble collapse exhibit very similar cost-function structures in the material-parameter space, with closely matching optimal parameters and corresponding 95\% confidence contours.
In particular, the elongated low-cost region extends from low viscosity and low shear modulus to high viscosity and high shear modulus, indicating offsetting effects between elasticity provided by a neo-Hookean spring and viscosity from a Newtonian dashpot in the first inertial collapse regime.
The close agreement between the material calibration landscapes and the optimal material parameters obtained from IMR and pIMR supports pIMR as an effective reduced-order alternative for approximating the material response up to the first collapse~\citep{zhu2025parsimonious}.

As a deterministic method, IMR exhibits comparatively large uncertainty when the fitting window is limited.
This uncertainty is progressively reduced as more data are incorporated, as shown for fitting windows extending to the second and third collapses.
The identified material properties change markedly between the first and second collapses, followed by only minor variations thereafter, suggesting distinct constitutive behavior at high and low strain rates and subsequent convergence of the inferred properties.
The converged values of shear modulus and viscosity are consistent with those previously reported in \citet{abeid2024experimental}.
By accounting for measurement noise during assimilation, En4D-Var exhibits consistently smaller variations across fitting-window lengths than standard IMR, indicating substantially reduced sensitivity to the choice of fitting window.
Though En4D-Var identifies optimal parameters with both larger shear modulus and viscosity for the data up to the first collapse, these values fall well within the elongated IMR cost-function structure. 
When the fitting window extends to the second and third collapses, En4D-Var results show favorable agreement with those calibrated by standard IMR, while exhibiting significantly reduced variations.
These results highlight the advantages of DA methods in accounting for measurement noise and uncertainty in LIC experiments for material characterization.
The characterization results for soft PAAm gels exhibit qualitatively similar trends and are presented in \cref{f:PA05003_contour}.


\begin{figure}[ht!]
    \centering
    \includegraphics[scale=1]{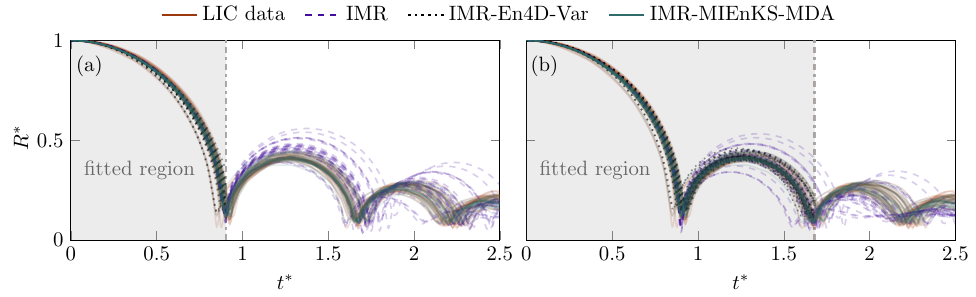}
    \caption{
    Predicted bubble dynamics in stiff PAAm gels at \SI{21}{\degreeCelsius}, using constitutive parameters calibrated from fitting windows extending to the (a) first and (b) second bubble collapses.
    The LIC measurements and the corresponding MIEnKS-MDA results are shown for comparison.
    }
    \label{f:PA_Rt}
\end{figure}

To directly compare the different IMR variants, \cref{f:PA_Rt} presents the predicted bubble dynamics in stiff PAAm gels at room temperature, based on the constitutive parameters identified by fitting windows of varying lengths in \cref{f:PA0503_contour}.
With the corresponding maximum and equilibrium radii for initialization, the bubble-radius trajectories predicted from the optimal IMR parameters remain nearly identical up to the first bubble collapse.
Later, they display a markedly larger spread than the experimental measurements.
By accounting for measurement noise and uncertainty, En4D-Var reduces this spread and improves the agreement between the data and the assimilated simulations, especially between the first and second collapses.
Nevertheless, some of the rebound bubble radii remain overpredicted.
By further accounting for the evolution of the constitutive properties, MIEnKS-MDA achieves accurate agreement with the experimental data, even beyond the third bubble collapse.
These results suggest that, within the constitutive model form of a neo-Hookean spring in parallel with a Newtonian dashpot, constant material properties can provide qualitative estimates but are insufficient to capture the full bubble dynamics.
Rather than changing the underlying constitutive form, allowing the effective constitutive properties to evolve meaningfully over time improves agreement with the measurements, reflecting the evolving strain-rate conditions throughout the cavitation process.


\begin{figure}[ht!]
    \centering
    \includegraphics[scale=1]{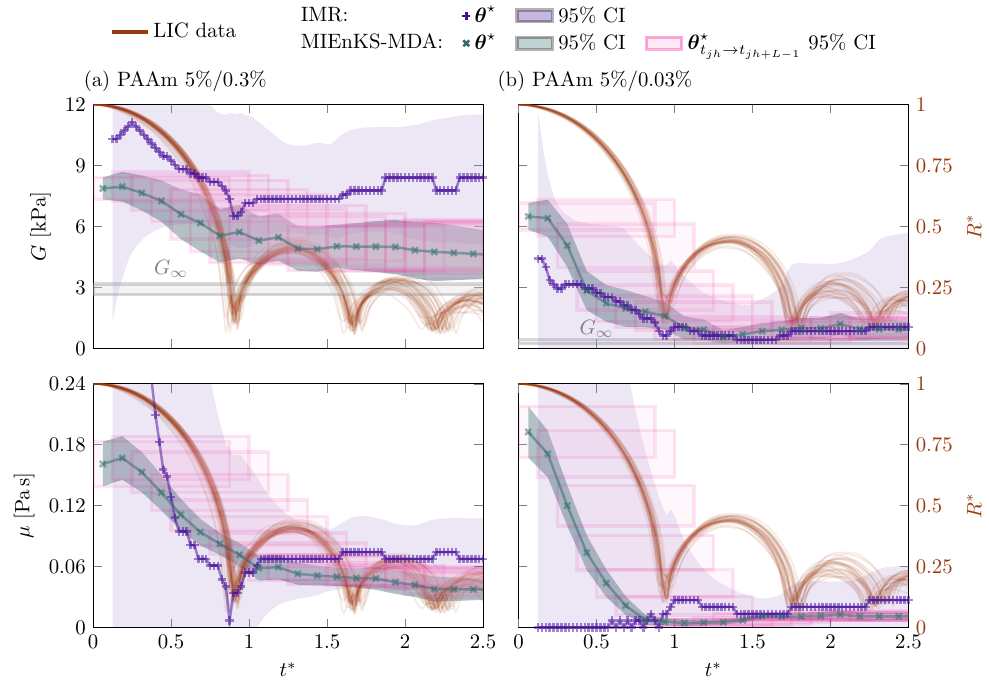}
    \caption{
        Constitutive properties throughout the cavitation process in PAAm gels at \SI{21}{\degreeCelsius}: (a) stiff PAAm; (b) soft PAAm.
        The IMR results represent the cumulative material response over the interval $[0,t^*]$ for $t^*\geq 0.125$, whereas MIEnKS-MDA provides locally averaged estimates of the transient behavior around $t^*$.
        The material properties inferred within each sliding assimilation window are also highlighted.
        The quasi-static shear modulus, $G_\infty$, corresponds to the elastic modulus reported in \citet{tse2010preparation}.
    }
    \label{f:PA property evolution_2D}
\end{figure}

\Cref{f:PA property evolution_2D} presents the constitutive properties identified throughout the cavitation process in PAAm gels at room temperature.
Both stiff and soft PAAm gels exhibit nearly constant shear-modulus responses after $t^* = 0.5$, while the viscosity decreases during the early stage of cavitation.
Subsequently, both quantities converge, indicating a steady constitutive response at low strain rates.
As expected, a higher bis-acrylamide concentration produces a stiffer material response~\citep{tse2010preparation}.
Thus, a larger effective shear modulus is obtained, and the viscosity shows comparatively weaker sensitivity to concentration~\citep{abeid2024experimental}.
The slight jumps in the IMR results arise solely from the discretization resolutions in $\Delta G$ and $\Delta \mu$.
For stiff PAAm, standard IMR yields relatively larger shear moduli and steady-state viscosity that characterize the overall behavior over the time window of $[0,t]$, whereas MIEnKS-MDA indicates lower effective transient values for both quantities.
For soft PAAm, MIEnKS-MDA yields a comparable shear modulus and a slightly higher steady-state viscosity than the optimal IMR estimates.
Moreover, for both the locally averaged estimates and the properties identified within each sliding assimilation window, the transient viscosity inferred by MIEnKS-MDA remains in favorable agreement with the IMR uncertainty envelope.
These results suggest that the mechanical response of PAAm gels under LIC initially decreases in both shear modulus and viscosity up to approximately the first bubble collapse, followed by a regime that can be effectively modeled by a nearly constant spring and dashpot in parallel.


\begin{figure}[ht!]
    \centering
    \includegraphics[scale=1]{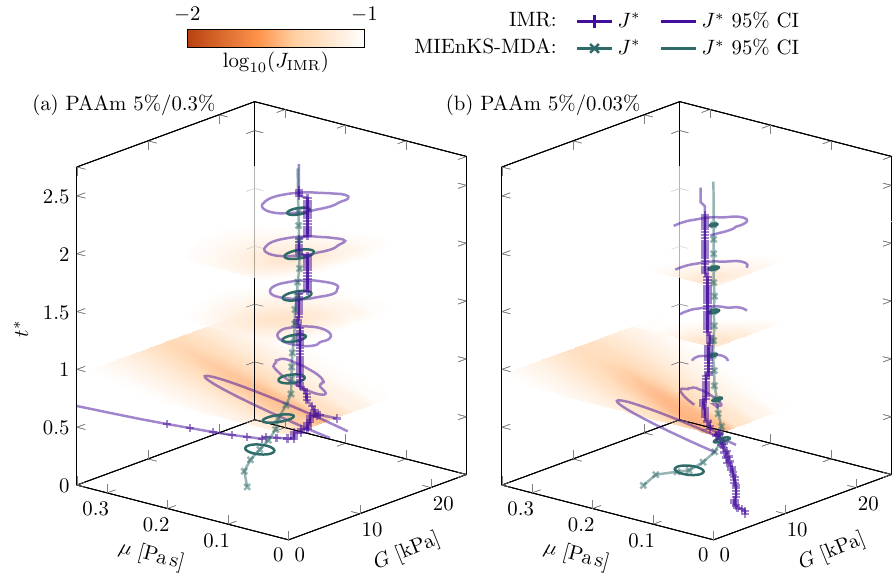}
    \caption{
     Constitutive-property trajectories in the $G$-$\mu$ parameter space throughout the cavitation process in PAAm gels at \SI{21}{\degreeCelsius}: (a) stiff PAAm; (b) soft PAAm.
     The cross-sectional contours correspond to the IMR cost-function values at different collapse times; see \cref{f:PA0503_contour} for stiff PAAm and \cref{f:PA05003_contour} for soft PAAm.
    }
    \label{f:PA property evolution}
\end{figure}

\Cref{f:PA property evolution} shows the temporal evolution of the constitutive properties in the material-parameter space for PAAm gels at \SI{21}{\degreeCelsius}.
At later stages of cavitation, both IMR and MIEnKS-MDA predict converged material properties together with their associated uncertainty intervals, indicating a steady constitutive response in both the overall behavior across the full time window and the locally averaged transient behavior.
Together with the trajectories in \cref{f:PA property evolution_2D}, the results show that MIEnKS-MDA initially infers higher shear modulus and viscosity than IMR, after which both quantities decrease as cavitation progresses and eventually approach lower or comparable steady-state values.
Throughout the cavitation process, IMR applied to the $57$ stiff and $30$ soft PAAm measurements exhibits large uncertainty bounds, making the optimal parameter estimates less reliable before and during the first bubble collapse ($t\lesssim 1$).
By contrast, MIEnKS-MDA yields comparatively consistent and well-bounded uncertainty estimates in this high-strain-rate regime and during subsequent evolution.
Overall, these results highlight the IMR-MIEnKS-MDA as a robust approach for calibrating constitutive properties from LIC measurements and resolving their temporal evolution within the NHKV constitutive model.

\subsection{Gelatin gels}\label{sec:gelatin}

\begin{figure}[ht!]
    \centering
    \includegraphics[scale=1]{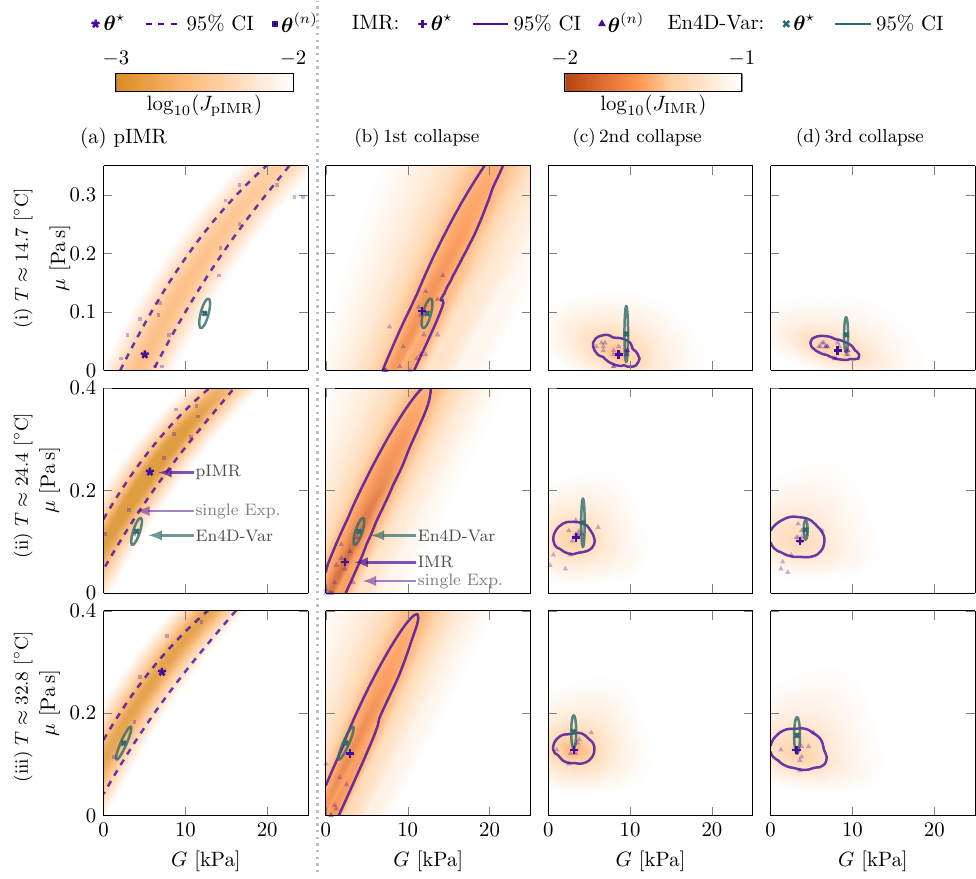}
    \caption{
        Calibrated material properties of gelatin gels at different temperatures, (i) $T\approx$~\SI{15}{\degreeCelsius}; (ii) $T\approx$~\SI{24}{\degreeCelsius}; and (iii) $T\approx$~\SI{33}{\degreeCelsius}.
        (a) shows the pIMR results and (b)--(d) show the standard IMR and IMR-En4D-Var results obtained with fitting windows extending to the 1st, 2nd, and 3rd bubble collapses, respectively.
        Background contours show the cost-function values in the material-parameter space for pIMR and IMR, along with the corresponding single-experiment estimates, $\vb*{\theta}^{(i)}$, the overall optimal parameters, $\vb*{\theta}^\star$, and the associated 95\% confidence contours.
        The mean and 95\% confidence region of the inferred multivariate Gaussian distributions are shown for En4D-Var.
    }
    \label{f:Gelatin_contour}
\end{figure}

\Cref{f:Gelatin_contour} compares the material properties calibrated for gelatin gels at different temperatures using fitting windows of varying lengths.
Unlike the nearly temperature-invariant results obtained for PAAm gels in \cref{f:PA0503_contour,f:PA05003_contour}, the inferred properties of gelatin display a clear temperature dependence.
Consistent with the bubble dynamics shown in \cref{f:Rt}~(c), the constitutive responses at $\SI{24.4}{\degreeCelsius}$ and $\SI{32.8}{\degreeCelsius}$ remain close to each other, while differing significantly from those at $\SI{14.7}{\degreeCelsius}$.
In particular, the calibrated shear modulus at $\SI{14.7}{\degreeCelsius}$ is significantly higher than at the two higher temperatures, which is consistent with the expected thermal softening associated with the gelatin melting.
The inferred properties thus change markedly between $\SI{14.7}{}$ and $\SI{24.4}{\degreeCelsius}$, consistent with proximity to the sol--gel transition.
We caution, however, against interpreting this change as an equilibrium transition temperature: the two warmer gelatin bins were measured during cooling from a fully melted state at $\SI{60}{\degreeCelsius}$ (\ref{App:Materials}), and because re-gelation of gelatin is not instantaneous on the \SIrange{15}{20}{\minute} measurement timescale, the network at these temperatures may be only partially re-formed.
The soft response above room temperature, therefore, reflects the combined effects of temperature and thermal history.


\begin{figure}[ht!]
    \centering
    \includegraphics[scale=1]{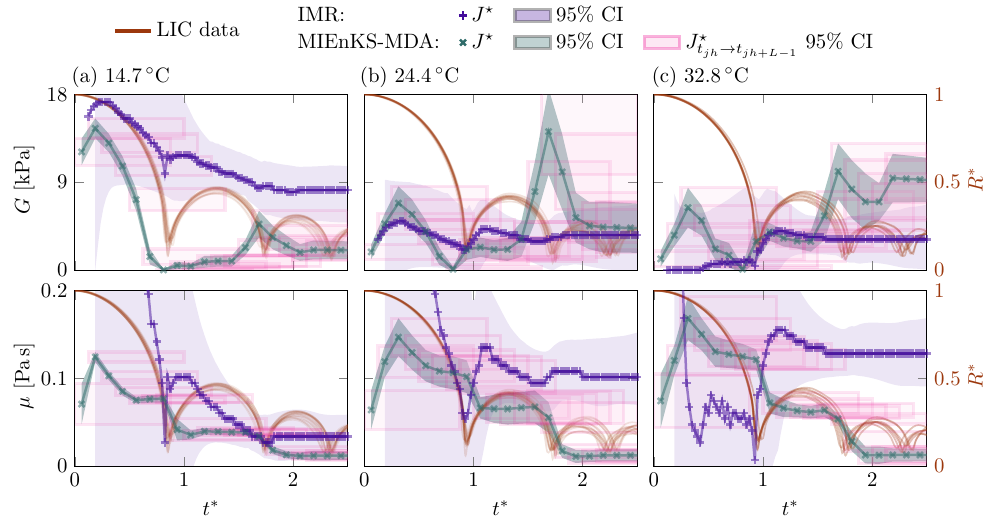}
    \caption{
        Temperature dependence of constitutive properties throughout the cavitation process in gelatin gels at approximately: (a) \SI{14.7}{\degreeCelsius}; (b) \SI{24.4}{\degreeCelsius}; and (c) \SI{32.8}{\degreeCelsius}.
        The IMR results represent the cumulative material response over the interval $[0,t^*]$ for $t^*\geq 0.125$, whereas MIEnKS-MDA provides locally averaged estimates of the transient behavior around $t^*$.
        The material properties inferred within each sliding assimilation window are also highlighted.
    }
    \label{f:Gelatin property evolution_2D}
\end{figure}

\Cref{f:Gelatin property evolution_2D} presents the evolution of the constitutive properties in gelatin gels and their coupling with the bubble dynamics at the three temperatures considered.
The viscous response is much less sensitive to temperature than the shear modulus.
Compared with the constitutive behavior of PAAm gels shown in \cref{f:PA property evolution_2D}, gelatin gels exhibit more pronounced variations during cavitation, indicating a more complex nonlinear viscoelastic response.
Standard IMR predicts increases in both shear modulus and viscosity at the first bubble collapse.
MIEnKS-MDA reveals qualitatively similar transient trends at all three temperatures: the apparent shear modulus generally decreases to near zero by the first bubble collapse, gradually recovers and rebounds near the second collapse, and then decreases again; the apparent viscosity initially rises to a maximum and then decreases in three stages, with the second and third stages occurring during the first two bubble collapses, while remaining nearly constant between collapses.
The near-vanishing apparent shear modulus around the first collapse should therefore not be interpreted as a true loss of elastic stiffness.
Instead, this behavior likely reflects the model-form error being absorbed by the effective NHKV parameters near the inertia-dominated minimum-radius state.
Similarly, the stepwise decrease in apparent viscosity is plausibly attributed to either progressive material damage during collapses or an additional rate-dependent response not fully represented by the NHKV model.
The low viscosity for $t^*\gtrsim 2$ suggests that the later stages of cavitation approach an inviscid-like response.
Interestingly, at \SI{14.7}{\degreeCelsius}, the shear modulus is initially much higher.
At sufficiently long times, it converges to a lower value than at the two higher temperatures, suggesting a more pronounced initial strain-stiffening effect when the material undergoes its maximum deformation.
This behavior may be further captured by an augmented neo-Hookean model with higher-order stretch-dependent terms~\cite{selvadurai2006deflections,yang2020extracting}.
The MIEnKS-MDA calibration at the two higher temperatures exhibits larger variances around the second collapse, making the subsequent inferred shear-modulus evolution less reliable.
Nevertheless, the trends up to $t^*\simeq 1.5$ remain qualitatively consistent with those obtained from standard IMR, revealing a similar transient softening during the first collapse.

\begin{figure}[ht!]
    \centering
    \includegraphics[scale=1]{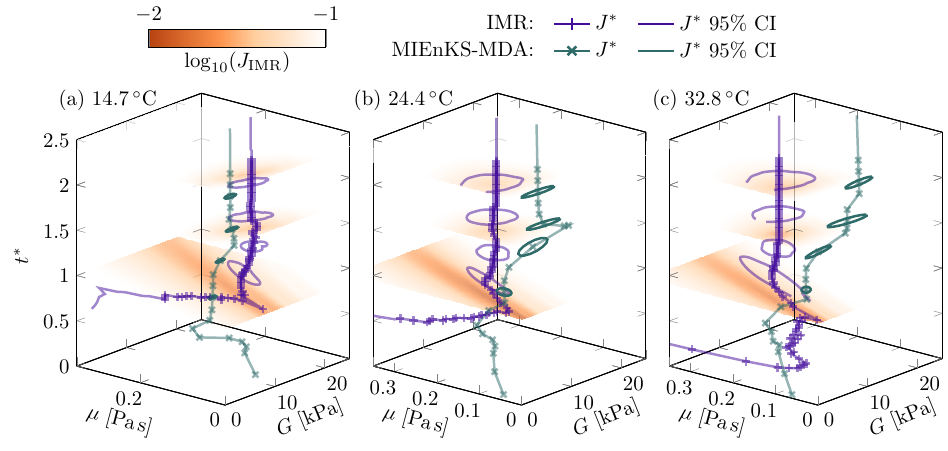}
    \caption{
     Constitutive-property trajectories in the $G$-$\mu$ parameter space throughout the cavitation process in gelatin gels at: (a) \SI{14.7}{\degreeCelsius}; (b) \SI{24.4}{\degreeCelsius}; and (c) \SI{32.8}{\degreeCelsius}.
     The cross-sectional contours correspond to the IMR cost-function values at different collapse times; see \cref{f:Gelatin_contour}.
    }
    \label{f:Gelatin property evolution}
\end{figure}

\Cref{f:Gelatin property evolution} further illustrates the temporal evolution of the constitutive properties of gelatin gels in the material-parameter space and reveals a clear temperature dependence.
Compared with the mechanical responses of PAAm gels shown in \cref{f:PA property evolution}, the gelatin gels exhibit much more pronounced differences between the evolution over the full time window and the locally averaged transient evolution inferred by the standard IMR and MIEnKS-MDA methods, respectively.
These results suggest that laser-induced cavitation gives rise to more complex transient constitutive behavior in gelatin.
Even so, the inferred material properties, together with their associated uncertainty intervals, still exhibit convergence.
Overall, the complex transient responses under high strain rates and the temperature sensitivity of gelatin suggest the need for more sophisticated constitutive models that can capture the material behavior across different regimes.

\section{Limitations of present work}\label{s:limits}

In the present implementation, the IMR-based characterization methods are demonstrated exclusively using the NHKV model, which consists of a neo-Hookean spring and a Newtonian dashpot connected in parallel.
We highlight that these IMR methods are not limited to this particular constitutive form and can be readily extended to a broad class of alternatives, including simpler formulations such as the linear Kelvin--Voigt-like~\citep{yang2005model,gaudron2015bubble} and the standard linear solid models~\citep{zener1949theory}, as well as more general nonlinear models such as the generalized quadratic Kelvin--Voigt~\citep{yang2020extracting,spratt2024numerical} and the standard nonlinear solid models~\citep{estrada2018high}.
Furthermore, the gelatin hydrogel experiments suggest that, within the prescribed constitutive structure of a spring and dashpot connected in parallel, assuming constant shear modulus and viscosity limits the model's ability to predict the full bubble dynamics.
This observation suggests the need to couple the NHKV model with rate- or state-dependent constitutive responses that can capture the mechanisms underlying the apparent evolution of elastic and viscous responses during inertial cavitation, such as finite-relaxation-spectrum effects.
Representative examples include the generalized Maxwell model~\cite{chockalingam2021probing}, which explicitly accounts for relaxation time scales, and the Bergstr\"{o}m--Boyce model~\cite{bergstrom1998constitutive,bergstrom2001constitutive}, which describes a nonlinear, rate-dependent stress response.
These alternative models involve different sets of constitutive parameters, and the reader is referred to \citet{warnez2015numerical,sanchez2025hierarchical} for broader overviews of the available constitutive model library.
The proposed MIEnKS-MDA framework can be extended accordingly to infer the evolution of these parameters.
Combined with Bayesian model selection~\citep{chu2025bayesian,sanchez2025hierarchical}, this rheological approach provides a convincing route to identify the most faithful constitutive model at different stages of cavitation.

Other than adjusting the ambient temperature according to our experimental conditions, we did not substantially modify the theoretical bubble dynamics model presented in \citet{estrada2018high}. 
Therefore, we implicitly accepted several underlying assumptions of that model; for example, the bubble reaches thermodynamic equilibrium at the end of the initial growth phase, and the surrounding viscoelastic material remains isothermal.
We note that these assumptions have been extensively validated for inertial cavitation in water~\cite{akhatov2001collapse, prosperetti1988nonlinear} but less so in soft hydrogels.
The pIMR technique is based on a reduced-order model of laser-induced cavitation at \qty{25}{\degreeCelsius}.
To delineate the effect of temperature on hydrogel viscoelasticity, further studies are needed to quantify the temperature-dependence of other factors present in the laser-induced cavitation process and, accordingly, modify the theoretical models used in the inverse characterization process.

\section{Conclusions}\label{s:conclusions}

Standard IMR and its data-assimilation variants infer a constant set of constitutive parameters whose values depend on the chosen fitting window.
We use this dependence to test the constant-parameter assumption directly: instead of reporting a single best-fit parameter set, we estimate how the effective parameters vary in time within a fixed model form, and identify where the constant-parameter model is insufficient to describe the dynamics.
The estimates are obtained from laser-induced cavitation (LIC) experiments using an IMR-based modified iterative ensemble Kalman smoother with multiple data assimilation (MIEnKS-MDA), a sliding-window rheometer that resolves the locally time-averaged constitutive response during cavitation.

Within the neo-Hookean Kelvin--Voigt (NHKV) constitutive model 
that consists of a neo-Hookean spring in parallel with a Newtonian dashpot, we obtain time-resolved estimates of the constitutive response of polyacrylamide (PAAm) hydrogels with different crosslinker concentrations.
The inferred shear modulus and viscosity generally decrease during cavitation and exhibit relatively weak temperature sensitivity.
In contrast, gelatin gels exhibit pronounced temperature dependence, with distinct property evolution trends at low and high temperatures.
In addition, both the apparent shear modulus and viscosity undergo substantial transient variations during the first two bubble collapses, highlighting the strongly coupled nature of bubble dynamics and material response of gelatin.
These time-resolved estimates are not introduced as arbitrary material parameters solely to improve the agreement with the measurements.
Instead, by allowing the effective NHKV parameters to vary within a prescribed constitutive structure, the method provides a diagnostic measure of how the constant-parameter NHKV assumption overconstrains the prediction of inertial cavitation dynamics.
The observed temporal variations, therefore, point to missing internal dynamics in the reduced constitutive description, such as finite relaxation spectra, history-dependent damage, or other evolving state variables, that may be needed to fully represent complex bubble--material interactions.

\section*{CRediT authorship contribution statement}

\textbf{TC}: Formal analysis, Methodology, Software, Investigation, Data Curation, Validation, Visualization, Writing – original draft, Writing – review $\&$ editing.
\textbf{JB}: Formal analysis, Methodology, Software, Investigation, Data Curation, Validation, Visualization, Writing – original draft, Writing – review $\&$ editing.
\textbf{ZZ}: Formal analysis, Methodology, Software, Investigation, Data Curation, Validation, Visualization, Writing – original draft, Writing – review $\&$ editing.
\textbf{JBE}: Conceptualization, Funding acquisition, Methodology, Project administration, Resources, Writing – review $\&$
editing.
\textbf{SHB}: Conceptualization, Funding acquisition, Methodology, Project administration, Resources, Supervision, Writing – review $\&$
editing.

\section*{Conflicts of interest}

There are no conflicts of interest to declare.

\section*{Data availability}

The experimental bubble-radius datasets and the code implementing the IMR-based characterization methods are available at \url{https://github.com/InertialMicrocavitationRheometry/Sliding_DA}.

\section*{Acknowledgments}

The authors acknowledge support from the U.S. Department of Defense, the Army Research Office under Grant No. W911NF-23-10324 (PMs Drs.\ Denise Ford and Robert Martin).
This work used PSC~Bridges2 and NCSA~Delta through allocation PHY210084 (PI Bryngelson) from the Advanced Cyberinfrastructure Coordination Ecosystem: Services $\&$ Support (ACCESS) program~\citep{boerner2023access}, which is supported by National Science Foundation grants $\#$2138259, $\#$2138286, $\#$2138307, $\#$2137603, and $\#$2138296.
JBE and ZZ acknowledge support from the National Science Foundation under grant $\#$2232426.

\setcounter{figure}{0}

\appendix

\section{IMR-based characterization}

\subsection{Parsimonious IMR (pIMR)}\label{App:pIMR}

Here, we briefly review the bubble-collapse analysis that leads to the pIMR method.
In the absence of (i) bubble content producing internal pressure, (ii) surface tension at the bubble wall, (iii) shear modulus in the surrounding material, and (iv) acoustic wave emission due to weak compressibility, the non-dimensional Keller--Miksis equation~\cref{eqn:keller-miksis} reduces to the form
\begin{equation}
    R \ddot{R} + \frac{3}{2} \dot{R}^{2} = -p_{\infty}^*.
\end{equation}
where $p_{\infty} = 1$. 
Lord~Rayleigh~\citep{rayleigh_1917} showed that, in this case, the collapsing bubble has a wall velocity
\begin{equation}
    \dot{R}^* = -\left( \frac{2}{3} \left(R^{*\,-3} - 1\right) \right)^{1/2} .
\end{equation} 
The total duration of the collapse process is then
\begin{equation}
    t_c^* = \int_{1}^{0} \frac{1}{\dot{R}^*} \dd{R^*} 
    = \frac{\sqrt{6\pi} \, \Gamma[11/6]}{5 \, \Gamma[4/3]} 
    \approx 0.9147,
\end{equation}
where $\Gamma[\cdot]$ is the gamma function and the resulting quantity is the Rayleigh collapse time, $t_{\rm RC}^*$.

\citet{zhu2025parsimonious} extended Lord Rayleigh's analysis to consider the physical factors that were assumed absent above. 
This formulation corresponds to the approximate form of the Keller--Miksis equation shown in \cref{eqn:approx-keller-miksis}, with correction factor $f$ serving as an additional, effective pressure term.
We then assume that the contribution of $f$ to the bubble collapse can be estimated according to its temporal average
\begin{equation}
    \overline{f} = \frac{1}{t_c} \int_{0}^{t_c} f \dd{t^*} 
    = \frac{1}{t_c} \int_{1}^{0} \frac{f}{\dot{R}^*} \dd{R^*}  
\end{equation}
such that 
\begin{equation}
    t_c^* \approx \int_{1}^{0}  -\left( \frac{2}{3}\left(1 - \overline{f}\right)\left(1 - R^{*\, -3}\right)   \right)^{-1/2} \dd{R^*} .
\end{equation}
Since time-averaging is a linear operator, the temporal average of the correction factor may be decomposed as $\overline{f} = \sum_{\alpha} \overline{f}_{\alpha}$, with $\alpha$ indexing different physical effects.
The specific forms of $\overline{f}_{\alpha}$ considered in this work are summarized below.

\paragraph{Bubble pressure}

The bubble pressure contribution to $f$ is
\begin{equation}
    \overline{f}_{\rm bc} = B \, p^*_{\rm go} R^{* \, 3\kappa}_0 + p^*_{\rm v}.
\end{equation}
Here, $\kappa$ is the ratio of the heat capacity at constant pressure, $C_{\rm P}$, to the heat capacity at constant volume, $C_{\rm V}$. 
$B$ is a scalar factor that is equal to 2.184 and 1.494, respectively, for the special cases of $\kappa = 1.4$ (isentropic) and $\kappa = 1$ (isothermal).
$p_{\rm v}^*$ and $p_{\rm go}^*$ are the steady-state pressures of the water vapor and non-vapor gas, respectively, constituting the bubble content.

\paragraph{Weak compressibility effect}

Compressibility is handled via the term
\begin{equation}
      \overline{f}_{\rm wc} = \frac{2 M_{\rm c}}{M_{\rm c} + \sqrt{M^2_{\rm c} + 4 t^{*2}_{\rm RC}}},
\end{equation}
where $M_{\rm c} = 1/c^*$ is the characteristic Mach number.

\paragraph{Surface tension}

Surface tension contributes to $\overline{f}$ as
\begin{equation}
    \overline{f}_{\rm We} = - \frac{\pi}{\sqrt{6} t^*_{\rm RC} \rm{We}}.
\end{equation}

\paragraph{Kelvin--Voigt viscoelasticity}
The viscoelastic term enters as
\begin{equation}
    \overline{f}_{\rm KV} = 
        \frac{4 \, \mathcal{C}}{2 \mathcal{C} + \sqrt{{\rm Re}^2 t^{*2}_{\rm RC} + 4 \, \mathcal{C}^2} } + 
        \frac{1}{\rm Ca} \left( \sqrt{\frac{2}{3}} \frac{R^*_0 \pi}{t_{\rm RC}^*} - \frac{5}{2} \right),
\end{equation}
where $R^*_0$ is the equilibrium radius of the bubble and $\mathcal{C}$ is a scalar factor that can be approximated as $\mathcal{C} \approx 0.4638 + 0.5639/{\rm Re} + 5.749/{\rm Re^2}$.
We refer the reader to \citet{zhu2025parsimonious} for the derivation and a comprehensive discussion of the above results.

\subsection{Ensemble-based four-dimensional variational~(En4D-Var) method}\label{App:En4dVar}

The optimization for the cost function in \cref{cost_en4dvar} is carried out using the form $\vb*{Q}_0  = \tilde{\vb*{Q}}_0+\vb*{Q}'_0\cdot\vb*{w}$ to restrict the solution to the subspace spanned by the scaled perturbation matrix around the initial ensembles using the correction coefficient $\vb*{w}$. 
This process is equivalent to finding the minimizer 
\begin{align}
    \vb*{w}_{\text{opt}} = \argmin_{\vb*{w}} J_w(\vb*{w})
\end{align}
for the cost function 
\begin{align}
    J_w(\vb*{w}) = 
        \frac{1}{2N_t} \sum_{k=1}^{N_t} \left\lVert\vb*{Y}_k^{\mathrm{D}}-\left<\vb*{Y}_k\right> - 
        \left<\vb*{Y}_k'\cdot \vb*{w}\right>\right\rVert^2_{\vb*{P}_k} + 
        \frac{1}{2}\vb*{w}^\top\vb*{w},
\end{align}
where the scaled output perturbation matrix takes the form of
\begin{align}
    \vb*{Y}'_k = \frac{1}{\sqrt{N_{\text{En}}-1}}
    \mqty[ \vb*{Y}_k^{(1)}-\left<\vb*{Y}_k\right>  & \cdots & \vb*{Y}_{k}^{(N_{\text{En}})}-\left<\vb*{Y}_k\right>].
\end{align}

In practice, we follow \citet{bocquet2014iterative} to seek the optimal correction coefficient $\vb*{w}_\text{opt}$ iteratively using a Gauss--Newton method,
\begin{align}
    \vb*{w}_{i+1}=\vb*{w}_i-\vb*{H}_i^{-1}  \nabla J_i(\vb*{w}_i),
\end{align}
where $i<N_\text{iter}$ is the iteration index, and $\vb*{H}$ and ${\nabla J}$ represent approximations of the Hessian and gradient of $J$.
They can be found with
\begin{align}
    \vb*{H}_i & = ({N_{\text{En}}-1})\vb*{I} 
    +\frac{1}{N_t}\sum_{k=1}^{N_t}  {\vb*{Y}'_k}^\top\vb*{P}_k^{-1}{\vb*{Y}'_k}, \\
    {\nabla J}_i & = -\frac{1}{N_t}\sum_{k=1}^{N_t}  {\vb*{Y}'_k}^\top\vb*{P}_k^{-1}\left(\vb*{Y}_k^{\mathrm{D}} - \left<\vb*{Y}_k\right>\right)  
    + ({N_{\text{En}}}-1)\vb*{w}_{i}.
\end{align}

To address the sampling error in ensemble methods due to finite ensemble size~\citep{whitaker2012evaluating}, we adopted the ``Relaxation Prior to Spread''~(RTPS) scheme by inflating the posterior distribution as 
\begin{align}
    \sigma_i = 
        {\sigma_i^{\mathrm{(post)}}} + a \left(
            {\sigma_i^{\mathrm{(prior)}}-\sigma_i^{\mathrm{(post)}}}
        \right), 
\end{align}
where $a = 0.7$ is an inflation parameter~\citep{spratt2021characterizing}. 

A similar restart strategy has been used in the restart-EnKF to address the dynamical systems with strong nonlinearity~\citep{zafari2005assessing, gu2007iterative, hendricks2008real}.
We apply En4D-Var to the data mean, and the measurement noise matrix $\vb*{P}_k$ at each time step is estimated from the variance of the measured radii across the $N_\mathrm{exp}$ experiments at that time step.
After obtaining the posterior ensemble, we restart the data assimilation process by drawing fresh samples from the inflated posterior distribution.
We repeat this process, and the final posterior distributions are obtained through $N_\mathrm{runs}=3$ complete cycles~\citep{chu2025bayesian}.
For reproducibility, \cref{tab:DA_params} summarizes the DA settings used throughout this work.

\begin{table}[ht!]
    \centering
    \caption{Data-assimilation hyperparameters used in this work.}\label{tab:DA_params}
    {\setlength{\tabcolsep}{9pt}
    \begin{tabular}{l c l}
        Parameter & Symbol & Value \\ \midrule
        Ensemble size & $N_\mathrm{En}$ & $48$ \\
        Restart cycles & $N_\mathrm{runs}$ & $3$ \\
        RTPS inflation factor & $a$ & $0.7$ \\
        Gauss--Newton iterations per cycle & $N_\mathrm{iter}$ & $5$ \\
        Sampling interval & $\Delta t^*$ & $0.025$ \\
        Window length and shift (MIEnKS-MDA) & $(L,h)$ & $(35,5)$ \\
        Prior mean $(G,\,\mu)$ &  & $(\SI{6}{\kilo\pascal},\,\SI{0.209}{\pascal\second})$ \\
        Prior standard deviation $(G,\,\mu)$ &  & $(\SI{2.8}{\kilo\pascal},\,\SI{0.088}{\pascal\second})$ \\
        Measurement-noise covariance & $\vb*{P}_k$ & sample variance across measurements \\
    \end{tabular}
    }
\end{table}


\subsection{Convergence of MIEnKS-MDA with assimilation-window length and shift}

\begin{figure}[ht!]
    \centering
    \includegraphics[scale=1]{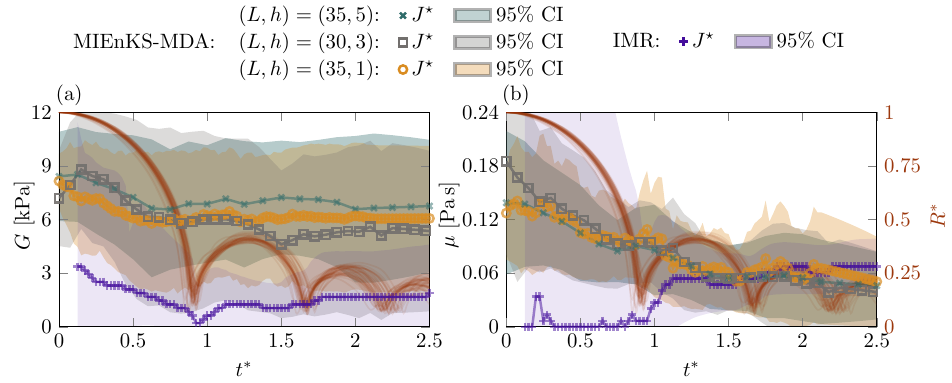}
    \caption{
    Convergence study of MIEnKS-MDA for constitutive-property calibration using different combinations of assimilation-window length and window-shift size, $(L,h)$.
   The shaded regions indicate the associated 95\% confidence intervals.
   Five measurements of the stiff PAAm gel at \SI{21}{\degreeCelsius} are used for demonstration.
    }
    \label{f:PA property evolution_convergence}
\end{figure}

\Cref{f:PA property evolution_convergence} shows a convergence study of MIEnKS-MDA for constitutive-property calibration using different combinations of assimilation-window length and window-shift size, $(L,h)$.
For demonstration, $5$ measurements of the stiff PAAm gel at \SI{21}{\degreeCelsius} are considered.
The results show that the current choice, $(L,h)=(35,5)$, produces qualitatively similar property estimates to those obtained by decreasing $L$ or $h$.
Over the full $120$ time steps, this choice requires only $18$ assimilation windows, compared with $31$ for $(L,h)=(30,3)$ and $87$ for $(L,h)=(35,1)$.
Thus, the selected setting provides comparable accuracy at substantially lower computational cost.
With only $5$ experiments, the standard IMR results exhibit noticeable discrepancies relative to those obtained with the full set of measurements in \cref{f:PA property evolution_2D}, indicating the limitations of IMR in handling uncertainty when the available data are sparse.

\section{Characterization of soft PAAm gels}\label{sec:PA05003}
\setcounter{figure}{0}

\begin{figure}[ht!]
    \centering
    \includegraphics[scale=1]{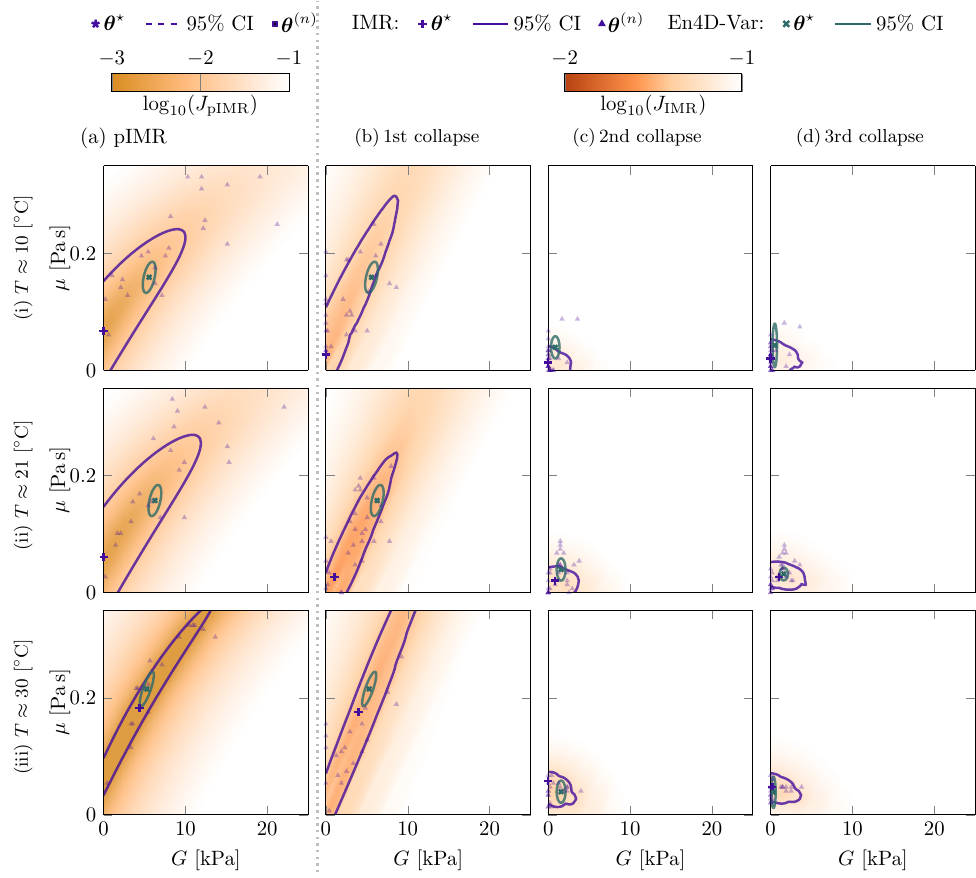}
    \caption{
        Calibrated material properties of soft PAAm gels at different temperatures, (i) $T\approx$~\SI{10}{\degreeCelsius}; (ii) $T\approx$~\SI{21}{\degreeCelsius}; and (iii) $T\approx$~\SI{30}{\degreeCelsius}.
        Panel (a) shows the pIMR results, while panels (b)–(d) show the IMR and IMR-En4D-Var results obtained with fitting windows extending to the 1st, 2nd, and 3rd bubble collapses, respectively.
        Background contours show the cost-function values in the material-parameter space for pIMR and IMR, along with the corresponding single-experiment estimates, $\vb*{\theta}^{(i)}$, the overall optimal parameters, $\vb*{\theta}^\star$, and the associated 95\% confidence contours.
        The mean and 95\% confidence region of the inferred multivariate Gaussian distributions are shown for En4D-Var.
    }
    \label{f:PA05003_contour}
\end{figure}

Analogous to \cref{f:PA0503_contour}, \cref{f:PA05003_contour} shows the material properties of soft PAAm gels calibrated using pIMR, IMR, and IMR-En4D-Var.
Similar to the stiff PAAm gels, the soft PAAm gels exhibit relatively weak temperature sensitivity.
At longer times, the converged shear modulus is substantially smaller than that of the stiff PAAm gels, consistent with the softer mechanical response of the material.
The qualitative differences among the IMR-based methods are also consistent with those observed in \cref{f:PA0503_contour}.

\section{Materials and Fabrication}\label{App:Materials}
\setcounter{figure}{0}

The 40\% acrylamide solution (CAS~79-06-1), the 2\% bis-acrylamide (CAS~110-26-9) solution, and the Tetramethylethylenediamine (TEMED; CAS 110-18-9) were sourced from Bio-Rad Laboratories (Hercules, CA, USA).
Ammonium persulfate (APS; CAS~7727-54-0) was purchased from MilliporeSigma (Burlington, MA, USA).
All PAAm gels were fabricated by creating a solution of 5\%w/w acrylamide and 0.3 or 0.03\%w/w bis-acrylamide based on the desired stiffness in deionized water, which constituted the remainder of the solution.
The mixture was briefly stirred on a vortex mixer.
After degassing under strong vacuum for \SI{15}{\minute}, 1\% of the total volume of a 10\%w/v APS solution in deionized water and 0.1\% of the total volume of TEMED were added to the mixture.
A vortex mixer on low speed was used to ensure a homogeneous solution.
Immediately afterward, \SI{3.5}{\milli\liter} of the mixture was transferred to \SI{4.5}{\milli\liter} polystyrene cuvettes (Fisher~Scientific, Waltham, MA, USA).

Knox Gelatin 5\%w/w in deionized water was heated on a hot plate set to \SI{100}{\degreeCelsius} while stirred at a moderate speed with a stir bar.
The beaker was covered to prevent solvent loss, even though the solution never reached boiling.
Once the gelatin had completely dissolved, \SI{3.5}{\milli\liter} of the solution was transferred into the same \SI{4.5}{\milli\liter} polystyrene cuvettes used for the PA. 

All cuvettes were sealed with Parafilm to prevent solvent evaporation.
All specimens were allowed to solidify for at least \SI{24}{\hour} before testing.
Specimens tested at non-ambient temperatures were stored in a refrigerator at approximately \SI{5}{\degreeCelsius} or a hot water bath at \SI{60}{\degreeCelsius} for at least \SI{12}{\hour} before testing. 
To characterize the temperature change when the specimens were reintroduced to an ambient environment, thermocouple measurements were taken with the probe placed near the center of the specimen.
At least two cooling and two warming experiments were run with every material tested. 
For cavitation testing, the exact time the specimens were removed from the refrigerator or hot water bath was recorded. 
The approximate temperatures of each cavitation were interpolated from an average time-temperature curve from the recorded thermocouple readings. 

\bibliographystyle{bibsty}
\bibliography{references}

\end{document}